\documentclass[twoside,11pt]{article}

\usepackage{amsmath, amsfonts, amssymb, amsthm}
\usepackage{mathrsfs,color}
\usepackage{bm,soul}
\usepackage{graphicx}
\usepackage{natbib}
\usepackage{subcaption}
\usepackage{float,xr,url}
\usepackage[left = 1in, right = 1in, top = 1in, bottom = 1in]{geometry}

\DeclareMathOperator*{\argmin}{arg\,min}

\DeclareMathOperator{\logit}{logit}

\DeclareMathOperator{\supp}{supp}

\def\trans{^{\T}}
\newcommand{\raisesymbol}[2]{\raisebox{\depth}{$#1#2$}}
\newcommand{\T}{{\mathpalette\raisesymbol\intercal}}

\DeclareMathOperator{\E}{\mathbb{E}}

\newcommand{\abs}[1]{\lvert{#1}\rvert}
\newcommand{\norm}[2][2]{\lVert{#2}\rVert_{#1}}

\def\independent{\protect\mathpalette{\protect\overlappingdouble}{\perp}}
\def\overlappingdouble#1#2{\mathrel{\rlap{$#1#2$}\mkern3mu{#1#2}}}

\newcommand{\RomanNum}[1]{\MakeUppercase{\romannumeral #1\relax}}

\let\wh\widehat
\let\wt\widetilde
\let\wb\overline

\newcommand{\DefineParameterVariants}[1]{
\expandafter\newcommand\expandafter{\csname b#1\endcsname}%
{\bm{\csname #1\endcsname}}
\expandafter\newcommand\expandafter{\csname b#1hat\endcsname}%
{\wh{\bm{\csname #1\endcsname}}}
\expandafter\newcommand\expandafter{\csname b#1til\endcsname}%
{\wt{\bm{\csname #1\endcsname}}}
\expandafter\newcommand\expandafter{\csname b#1bar\endcsname}%
{\wb{\bm{\csname #1\endcsname}}}
\expandafter\newcommand\expandafter{\csname b#1ast\endcsname}%
{{\bm{\csname #1\endcsname}}^\ast}
\expandafter\newcommand\expandafter{\csname #1hat\endcsname}%
{\wh{\csname #1\endcsname}}
\expandafter\newcommand\expandafter{\csname #1til\endcsname}%
{\wt{\csname #1\endcsname}}
\expandafter\newcommand\expandafter{\csname #1bar\endcsname}%
{\wb{\csname #1\endcsname}}
\expandafter\newcommand\expandafter{\csname #1ast\endcsname}%
{{\csname #1\endcsname}^\ast}
}

\DefineParameterVariants{alpha}
\DefineParameterVariants{omega}
\DefineParameterVariants{tau}
\DefineParameterVariants{xi}

\DefineParameterVariants{beta}
\DefineParameterVariants{delta}
\DefineParameterVariants{rho}
\DefineParameterVariants{zeta}
\DefineParameterVariants{gamma}
\DefineParameterVariants{theta}

\DefineParameterVariants{Sigma}
\newcommand{\kappaast}{\kappa^\ast}

\newcommand{\init}{{\text{init}}}

\newcommand{\Deltaalpha}{\Delta_{\balpha}}

\newcommand\blfootnote[1]{%
  \begingroup
  \renewcommand\thefootnote{}\footnote{#1}%
  \addtocounter{footnote}{-1}%
  \endgroup
}

\newcommand{\DefineNumericVariants}[1]{
\expandafter\newcommand\expandafter{\csname b#1\endcsname}%
{\bm{#1}}
\expandafter\newcommand\expandafter{\csname s#1\endcsname}%
{\mathscr{#1}}
\expandafter\newcommand\expandafter{\csname bl#1\endcsname}%
{\mathbb{#1}}
}

\DefineNumericVariants{X}
\DefineNumericVariants{Z}

\DefineNumericVariants{U}
\DefineNumericVariants{V}

\DefineNumericVariants{H}
\DefineNumericVariants{G}
\DefineNumericVariants{K}

\DefineNumericVariants{L}
\DefineNumericVariants{E}

\DefineNumericVariants{T}
\DefineNumericVariants{M}

\DefineNumericVariants{a}
\DefineNumericVariants{d}
\DefineNumericVariants{u}
\DefineNumericVariants{v}

\DefineNumericVariants{R}
\DefineNumericVariants{C}

\newcommand{\bZalpha}{\bZ_{\balpha}}
\newcommand{\bZalphahat}{\bZ_{\balphahat}}
\newcommand{\bZalphaast}{\bZ_{\balphaast}}

\definecolor{jcolor}{RGB}{041,122,000}
\definecolor{darkred}{RGB}{100,000,000}
\definecolor{purple}{RGB}{100,000,150}
\def\purple{\color{black}}

\usepackage{mwe}

\newcommand{\commentout}[1]{}

\newcommand{\DefineSetVariants}[1]{
\expandafter\newcommand\expandafter{\csname c#1\endcsname}%
{{\mathcal{#1}}}
\expandafter\newcommand\expandafter{\csname c#1ast\endcsname}%
{{\mathcal{#1}^\ast}}
\expandafter\newcommand\expandafter{\csname c#1hat\endcsname}%
{{\wh{\mathcal{#1}}}}
\expandafter\newcommand\expandafter{\csname c#1bar\endcsname}%
{{\wb{\mathcal{#1}}}}
}

\newcommand{\DefineSetComplementVariants}[1]{
\expandafter\newcommand\expandafter{\csname c#1comp\endcsname}%
{{\cP \setminus \mathcal{#1}}}
\expandafter\newcommand\expandafter{\csname c#1astcomp\endcsname}%
{{\cP \setminus \mathcal{#1}^{\ast}}}
\expandafter\newcommand\expandafter{\csname c#1hatcomp\endcsname}%
{{\cP \setminus \wh{\mathcal{#1}}}}
}

\DefineSetVariants{P}  
\DefineSetVariants{A}  
\DefineSetVariants{Q}  
\DefineSetVariants{S}  
\DefineSetVariants{B}  
\DefineSetVariants{J}  
\DefineSetVariants{M}  

\def\subone{_{\scriptscriptstyle 1}}

\DefineSetComplementVariants{A}

\def\nnhalf{n^{-1/2}}
\def\ninv{n^{-1}}

\newcommand{\cSplus}{{\cS_{\text{\texttt{+}}}}}
\newcommand{\cSminus}{{\cS_{\text{\texttt{-}}}}}

\newcommand{\cSplusast}{{\cS_{\text{\texttt{+}}}^{\ast}}}
\newcommand{\cSminusast}{{\cS_{\text{\texttt{-}}}^{\ast}}}

\newcommand{\cSplusbar}{{\wb{\cS}_{\text{\texttt{+}}}}}
\newcommand{\cSminusbar}{{\wb{\cS}_{\text{\texttt{-}}}}}

\newcommand{\zero}{\bm{0}}

\newtheorem{case}{Case}

\newcommand{\ModelY}{\ensuremath{\mathcal{M}_Y}}
\newcommand{\ModelS}{\ensuremath{\mathcal{M}_S}}
\newcommand{\Cprior}{{\ensuremath{\mathcal{C}^{\text{prior}}}}}
\newcommand{\SSprior}{{\ensuremath{\text{SS}^{\text{prior}}}}}
\newcommand{\SSulasso}{{\ensuremath{\text{SS}^{\text{ULASSO}}}}}

\newtheorem{proposition}{Proposition}
\newtheorem{theorem}{Theorem}

\theoremstyle{definition}
\newtheorem{remark}{Remark}
\newtheorem{rexample}{Data Example}
\def\bW{\mathbf{W}}

\def\trans{^{\T}}

\begin{document}

    \title{Prior Adaptive Semi-supervised Learning with Application to EHR Phenotyping}
    \date{}
          \author{Yichi Zhang$^{1*}$, \blfootnote{Zhang and Liu contributed equally to this work.} Molei Liu$^{2*}$, Matey Neykov$^{3}$ and Tianxi Cai$^{2}$ }
    \maketitle
    \thispagestyle{empty}

\footnotetext[1]{Department of Computer Science and Statistics, University of Rhode Island.}
\footnotetext[2]{Department of Biostatistics, Harvard T.H. Chan School of Public Health.}
\footnotetext[3]{Department of Statistics and Data Science, Carnegie Mellon University.}

\begin{abstract}

\noindent Electronic Health Records (EHR) data, a rich source for biomedical research,
have been successfully used to gain novel insight into a wide range of diseases. Despite its potential, EHR is currently underutilized for discovery research due to it's
major limitation in the lack of precise phenotype information. To overcome such
difficulties, recent efforts have been devoted to developing supervised algorithms to
accurately predict phenotypes based on relatively small training datasets with gold
standard labels extracted via chart review. However, supervised methods typically
require a sizable training set to yield generalizable algorithms especially when the
number of candidate features, $p$, is large. In this paper, we propose a semi-supervised
(SS) EHR phenotyping method that borrows information from both a small labeled
data where both the label $Y$ and the feature set $\bX$ are observed and a much larger
unlabeled data with observations on X only as well as a surrogate variable $S$ that
is predictive of $Y$ and available for all patients, under a high dimensional setting.
Under a {\em working} prior assumption that $S$ is related to $\bX$ only through $Y$ and allowing it to hold {\em approximately}, we propose a prior adaptive semi-supervised (PASS)
estimator that adaptively incorporates the prior knowledge by shrinking the estimator towards a direction derived under the prior. We derive asymptotic theory for
the proposed estimator and {\purple justify its efficiency and robustness to prior information of poor quality}. We also demonstrate its superiority over existing estimators under various scenarios via
simulation studies. The proposed method is applied to {\purple three EHR phenotyping studies at a large tertiary hospital system and outperforms existing methods in the real application.}

\vspace{0.15in}
\end{abstract}

\section{Introduction}

Electronic Health Records (EHRs) provide a large and rich data source for biomedical research aiming to further our understanding of disease progression and treatment response. EHR data has been successfully used to gain novel insights into a wide range of diseases, with examples including diabetes \citep{brownstein2010}, rheumatoid arthritis \citep{liao2014}, inflammatory bowl disease \citep{ananthakrishnan2014}, and autism \citep{doshi2014}. EHR is also a powerful discovery tool for identifying novel associations between genomic markers and multiple phenotypes through analyses such as phenome-wide association studies \citep{denny2010,kohane2011,wilke2011,cai_phewas2018}.

Despite its potential, ensuring unbiased and powerful biomedical studies using EHR is challenging because EHR was primarily designed for patient care, billing, and record keeping. 
Extracting precise phenotype information for an individual patient requires manual medical chart review, an expensive process that is not scalable for research studies. 
To overcome such difficulties, recent efforts including those from Informatics for Integrating Biology and the Bedside (i2b2)  \citep[e.g.]{liao2015a,yu2015} and the Electronic Medical Records and Genomics (eMERGE) network \citep{newton2013,gottesman2013} have been devoted to developing phenotyping algorithms to predict disease status using relatively small training datasets with gold standard labels extracted 
via chart review. 

Various approaches to EHR phenotyping have been proposed. Supervised machine learning methods have been shown to achieve robust performance across disease phenotypes and 
EHR systems \citep{carroll2012,liao2015a}. However, supervised methods typically require a sizable training set to yield generalizable algorithms especially when  
the candidate features, denoted by $\bX$, is of high dimension. One approach to overcome the high dimensionality is to consider unsupervised methods. Unfortunately, 
standard unsupervised methods such as clustering are likely to fail when the dimension of $\bX$ is large but a majority of the features are unrelated to the phenotype of interest but 
possibly predictive of some other underlying subgroups. Recently, unsupervised methods based on ``silver standard labels" have been proposed. {\purple These methods leverage a surrogate outcome $S$,  highly predictive of the true phenotype status $Y$, such as the count of International Classification of Diseases (ICD) billing codes of the disease,  to train the phenotyping algorithm against the features $\bX$. Specifically, \cite{halpern2016electronic} and \cite{zhang2020maximum} utilized anchor variables with high positive predictive value as the surrogate $S$ to estimate $Y\mid \bX$ under the conditional independence assumption $S\independent\bX\mid Y$.} \cite{agarwal2016} trained penalized logistic regression on $S\sim\bX$ for phenotyping of $Y$ against $\bX$.  \cite{chakrabortty2017} provided theoretical justification for this strategy. {\purple They showed that a regularized estimator constructed from an unlabeled subset consisting of those with extreme values of $S$ can be used to make inference about the direction of $\bbeta$ under single index models $S \sim f(\balpha\trans\bX,\epsilon)$ and $Y \sim g(\bbeta\trans\bX)$. Their method relies on the similarity between the directions of $\balpha$ and $\bbeta$ to make efficient estimation. However, it is not robust to poor surrogacy resulted from violation of such assumptions. Furthermore, their method cannot be directly used to predict $Y$ using both $S$ and $\bX$. In addition, these unsupervised approaches typically cannot accurately recover the scale of $P(Y=1\mid S,\bX)$.} 

{\purple A number of semi-supervised or weakly supervised deep learning procedures have also been proposed recently and shown to attain better performance than the supervised counterparts. For example, \cite{ratner2017snorkel} proposed a weakly supervised approach that trains a deep model with imperfect labels generated from user--specified label functions from sources such as patterns, heuristics, and external knowledge bases. \cite{wang2018deep} developed a framework for weak supervision from multiple sources by composing probabilistic logic with deep learning. \cite{mcdermott2018semi} designed a semi-supervised cycle Wasserstein regression generative adversarial networks (CWR-GAN) approach using adversarial signals to learn from unlabelled samples and improve prediction performance in scarcity of gold-standard labels. However, it remains unclear when and how the surrogate features along with the unlabeled data can improve the prediction performance of these deep models, due to their complex architectures.}


In this paper, we propose an {\purple semi-supervised (SS)} method for estimating $Y \mid \bW=(S, \bX\trans)\trans$ that borrows information from both a small  labeled data with $n$ realizations of $(Y, \bW\trans)\trans$ and a much larger unlabeled data with $N$ observations on $\bW$, under a high dimensional setting with $N \gg p \gg n$. 
We consider a logistic phenotype model for $Y \mid S, \bX$, a single index model (SIM) for $S \mid \bX$, as well as a {\em working} prior assumption that $S$ is independent of $\bX$ given $Y$. We obtain the estimator through regularization with penalty functions reflecting the prior knowledge. When the prior assumption holds exactly, we show that the unlabeled data can naturally be used to assist in the estimation of the phenotype model. 
Allowing the prior assumption to hold approximately {\purple or to be highly violated}, our prior adaptive semi-supervised (PASS) estimator 
adaptively incorporate the prior knowledge by shrinking the estimator towards a direction derived under the
prior. 

The proposed PASS estimator is similar to the prior LASSO (pLASSO) procedure of \citet{jiang2016}  in spirit in that 
both approaches aim to incorporate prior information into the $\ell_1$ penalized estimator in a 
high-dimensional setting. The differences are, nevertheless, substantial and clear.
\citet{jiang2016} assumed that the prior information
was summarized into prediction values and contributed
to the likelihood term. In contrast,
we use prior information to guide the shrinkage
and put them into the penalty term.
In this sense, PASS and pLASSO complement each other to some extent.
However, as shown in both theory and simulations,
putting prior information into the likelihood term
tends to lead to the ``take it or leave it'' phenomenon:
the usefulness of the prior information is determined 
based on the overall effect of all predictors.
On the other hand, by putting prior information into the penalty term, the PASS 
approach provides more flexible control: the data is able to scrutinize the individual effect of each predictor.
This gained flexibility can result in improved theoretical and numerical performances.

The rest of this paper is organized as follows. We discuss the motivation, an important special scenario and the general methodology in Section~2.
We analyze the theoretical properties of the proposed approach
in Section~3,
and access its finite sample performance via simulation studies
in Section~4.
Furthermore, we illustrate the practical value of the proposed
approach on three real EHR datasets in Section~5.
Finally, we conclude this paper with some discussions and extensions 
in Section~6. All technical proofs and additional numerical results are given in the Supplementary Materials.

\def\bW{\mathbf{W}}

\section{Methodology}

\subsection{Setup}

We assume that the underlying data consists of $N$ independent and identically distributed (i.i.d.) observations
$\{(Y_i, S_i, \bX_i^\T)^\T = (Y_i, \bW_i\trans)\trans, i = 1, \dots, N\}$,
where $Y_i$ is a binary indicator of the disease status of the $i$th patient,
$S_i$ is a scalar surrogate variable that is reasonably predictive of $Y_i$ chosen via domain knowledge, 
and $\bX_i$ is a $p$-dimensional feature vector. Examples of $S_i$ includes the total count of ICD codes or mentions for the disease of interest in clinical notes extracted via natural language processing (NLP). Candidate features $\bX$ may include the ICD9 code counts for competing diagnosis, lab results, as well as NLP mentions of relevant signs/symptoms, medications and procedures. We may also include various transformations of original features in $\bX$ to account for non-linear effects. 
While $\{\bW_i, i = 1, ..., N\}$ is fully observed, $Y_i$ is only observed for a random subset of $n$ patients.
Hence the observed data are $\sL \cup \sU$, 
where without loss of generality, 
the first $n$ observations are assumed fully observed as 
$\sL = \{(Y_i, \bW_i^\T)^\T, i = 1, \dots, n\}$, and 
the rest constitute the unlabeled data as
$\sU = \{\bW_i, i = n + 1, \dots, N\}$.

Throughout, for a $d$-dimensional vector
$\bu$, the $\ell_q$-norm of $\bv$
is $\norm[q]{\bv} = (\sum_{j=1}^d \abs{v_j}^q)^{1/q}$.
The $\ell_\infty$-norm of $\bv$ is
$\norm[\infty]{\bv} = \max_{1 \leq j \leq d} \abs{v_j}$. The support of $\bv$ is $\supp(\bv) = \{j: v_j \neq 0\}$. 
If $\cJ$ is a subset of $\{1, \dots, p\}$,
then $\bv_{\cJ}$ denotes a $d$-dimensional vector
whose $j$th element is $v_j 1_{j \in \cJ}$,
and $1_{B}$ is the indicator function for set $B$.
The independence between random variables/vectors $\bU$ and $\bV$
is written as $\bU \independent \bV$.
We also denote the negative log-likelihood function 
associated with the logistic model with $\ell(y, \eta) = -y \eta + \log(1 + e^\eta)$.

\def\bvartheta{\boldsymbol{\vartheta}}
\def\bWvec{\vec{\bW}}
\def\bw{\mathbf{w}}

\subsection{Model Assumptions}\label{sec:method:assume}
\label{sec-model}
To predict $Y$ using $\bW=(S,\bX\trans)\trans$, we assume
\begin{equation}
\label{Logistic}
 \Pr(Y = 1 \mid \bW) = \pi(\zeta_0 + S \gamma_0 + \bX^\T \bbeta_0) = \pi(\bvartheta_0\trans\bWvec)\quad
\mbox{with}\quad \bvartheta_0 = (\zeta_0, \gamma_0, \bbeta_0\trans)\trans, 
\tag{\ModelY}
\end{equation}
where for any vector $\bw$, $\vec\bw = (1, \bw\trans)\trans$, and $\pi(t) = e^t / (1 + e^t)$. 
{\purple To leverage the data in $\sU$, 
we further assume a single index model (SIM) for $S \mid \bX$, i.e. there exists $\balpha_0\in\mathbb{R}^p$ such that
\begin{equation}
\label{SIM}
S = f(\bX^\T \balpha_0, \epsilon),  
~\text{with some } \epsilon \independent \bX
\text{ and } f \text{ satisfying } \E \{f^2(\bX^\T \balpha_0, \epsilon)\}<\infty,
\tag{\ModelS}
\end{equation}
where $\bX^\T \balpha_0$ is a single linear combination of the features $\bX$ and $f$ is an unknown link function.}
Here $\zeta_0$, $\gamma_0$, $\bbeta_0$ and $\balpha_0$
are parameters to be estimated {\purple where only the direction of $\balpha_0$ is  identifiable and its norm does not affect our construction introduced below.} If $\balpha_0$ and $\bbeta_0$ are similar in certain ways, one would expect that the unlabeled data $\sU$ 
may be used to improve upon the 
standard supervised estimator for $\bbeta_0$ using $\sL$ alone. 
For example, if $S$ is a noisy representation of $Y$ with random measurement error, 
then it is reasonable and {\purple common in the EHR literature \citep[e.g.]{hong2019semi,zhang2020maximum}} to assume
\begin{equation}
\bX \independent S \mid Y. \label{ass-ind} \tag{\Cprior}
\end{equation}
Under  (\ref{ass-ind}), we have Proposition \ref{thm:perfect-surrogate}
with proof  given in Supplementary Materials.

\begin{proposition}
\label{thm:perfect-surrogate}
Under (\ref{Logistic}), (\ref{SIM}), (\ref{ass-ind}), and assume {\purple $\E(\bX\bX^\T)$ is positive--definite}, and it holds that: (C1) for any two vectors $\ba_1, \ba_2$, 
$\E(\bX^\T \ba_2 \mid \bX^\T \ba_1)$ is linear in $\bX^\T \ba_1$. {\purple Then there exist scalars $k_1,k_2\in\mathbb{R}$ such that} $\balpha_0=k_1\bbeta_0$ and $\balpha^* = k_2 \bbeta_0$ where 
$$(\tau^*,\balpha^*) = \argmin_{\tau,\balpha}
{\E (S - \tau - \bX^\T \balpha)^2}.$$
\end{proposition}
{\purple 
\begin{remark}
{Condition~(C1) holds for elliptical distributions including multivariate normal. By \cite{diaconis1984} and \cite{hall1993}, this assumption tends to hold for non-elliptical design when the dimensionality is high. Specifically, one can show that under mild regularity conditions, for two projection vectors $\ba_1$ and $\ba_2$ uniformly randomly drawn from $\mathbb{S}^{p-1}=\{\bv\in\mathbb{R}^{p-1}:\|\bv\|_2=1\}$, the pair $(\bX^\T \ba_2,\bX^\T \ba_1)$ weakly converges to a bivariate normal distribution with high probability, and thus $\E(\bX^\T \ba_2 \mid \bX^\T \ba_1)$ is at least approximately linear in $\bX^\T \ba_1$; see Theorem 1.1 of \cite{diaconis1984} and equation (1.9) of \cite{hall1993}.}
\end{remark}
}

{\purple Proposition~\ref{thm:perfect-surrogate} hinges on the main result of \cite{li1989} that when the features $\bX$ satisfy (C1), the direction of the coefficients of a SIM could be estimated using least squares regression for the response against $\bX$.} It suggests that 
 $\sU$ can greatly improve the estimation of $\bbeta_0$ under \eqref{ass-ind} 
because the phenotype model \eqref{Logistic} 
may be rewritten as
$\logit \Pr (Y = 1 | \bW) = 
\zeta + S \gamma + \rho \bX^\T \balpha$ 
for some $\rho$. 
Under this model, a simple SS estimator for 
$\zeta$, $\gamma$ and $\bbeta$ in \eqref{Logistic} can be obtained as
$\zetahat$, $\gammahat$ and $\rhohat \balphahat$, where  {\purple 
\[
(\widehat\zeta, \widehat\gamma, \widehat\rho)^\T=\argmin_{\zeta, \gamma, \rho} \sum_{i=1}^n \ell(Y_i, \zeta + \gamma S_i + \rho \bX_i^\T \balphahat),\quad(\tauhat, \balphahat^\T)^\T=\argmin_{\tau,\balpha}\sum_{i=1}^N (S_i - \tau - \bX \balpha)^2.
\]}
By doing so, the direction of the high dimensional vector $\bbeta$
is estimated based on the entire data $\sL \cup \sU$,
and only the parameters $(\zeta, \gamma, \rho)^\T$ are estimated 
using the small labeled data $\sL$. 
Hereafter we shall refer to this SS estimator derived under \eqref{ass-ind} as $\SSprior$.

Nevertheless, $\SSprior$ is only valid when \eqref{ass-ind} and (C1) holds exactly. 
Our goal is to develop a more robust SS estimator under \eqref{Logistic} and \eqref{SIM} that can efficiently exploit $\sU$ when \eqref{ass-ind} and (C1) may only hold approximately. 
In this more general setting, a desirable SS estimator should improve
upon the standard supervised estimator when the directions 
of $\balpha_0$ and $\bbeta_0$ are similar in their magnitude and/or support. In addition, it should perform similarly to the supervised estimator when the two directions are not close. We shall now detail our PASS estimation procedure 
which automatically adapts to different cases as reflected in the observed data.

\subsection{Prior Adaptive Semi-Supervised (PASS) Estimator}

With $\sL$ only, a supervised estimator for $\bbeta$ can be obtained via the  standard $\ell_1$-penalized regression: 
\begin{equation}
\label{eq:supervised-beta}
\breve\bvartheta=(\breve\zeta, \breve\gamma, \breve\bbeta\trans)\trans = \argmin_{\bvartheta} 
\frac{1}{n} \sum_{i=1}^n 
\ell(Y_i, \bvartheta\trans \bWvec_i)
+ \lambda \norm[1]{\bbeta}. 
\end{equation}
With properly chosen $\lambda$, the consistency and rate of convergence for $\breve\bvartheta$ 
has been established \citep{vandegeer2008}. To improve the estimation of  $\bbeta$ through leveraging $\sU$, we note that when \eqref{ass-ind} holds approximately, the magnitude of $\bbeta_0 - \rho \balpha_0$ is small for some $\rho$, and the support of $\bbeta_0 - \rho \balpha_0$ is of small size as well. 

To incorporate such prior belief on the relationship between $\balpha_0$ and $\bbeta_0$, we construct the penalty term
\[
\min_\rho \{\lambda_1 \norm[1]{(\bbeta - \rho \balpha_0)_{\cA_0}}
+ \lambda_2 \norm[1]{(\bbeta - \rho \balpha_0)_{\cA_0^c}}\},
\]
where $\cA_0 = \supp(\balpha_0)$, and 
$\lambda_1, \lambda_2 > 0$ are tuning parameters.
Since $(\balpha_0)_{\cA_0^c} = \bm{0}$, the penalty term
is equivalent to
\begin{equation}
\label{eq:penalty}
\lambda_1 \{ \min_\rho \norm[1]{(\bbeta - \rho \balpha_0)_{\cA_0}} \}
+ \lambda_2 \norm[1]{\bbeta_{\cA_0^c}}.
\end{equation}
The first term in the penalty measures how far
$\bbeta$ is from the closest vector along the $\balpha_0$ direction,
and hence encourages smaller magnitude of
$\bbeta - \rho \balpha_0$. 
The second term shrinks $\bbeta_{\cA_0^c}$ towards $\bm{0}$,
which reflects our prior that predictors irrelevant to $S$
are likely to be irrelevant to $Y$ as well.
The tuning parameters $\lambda_1, \lambda_2$
control the strength of the belief imposed.
When they are sufficiently large,
$\bbeta$ will be forced to be a multiple of $\balpha_0$
and thus it ends up with the same estimator as in 
the case where \eqref{ass-ind} holds.

Since we have $N\gg p$ samples to estimate $\balpha_0$, we use the adaptive LASSO (ALASSO) penalized least square estimator $\balphahat$
\citep{zou2006,zou2009}, where
\[
\tauhat, \balphahat = \argmin_{\tau, \balpha} 
\frac{1}{N} \sum_{i=1}^N \left( S_i - \tau - \bX_i^\T \balpha \right)^2 
+ \mu \sum_{j=1}^p \omegahat_j \abs{\alpha_j},
\]
where $\omegahat_j = \abs{\alphahat_{\init, j}}^{-\nu}$ for some constant $\nu>0$, $\balphahat_\init = (\alphahat_{\init,1}, ...,\alphahat_{\init,p})\trans$,
\[
\tauhat_\init, \balphahat_\init =  \argmin_{\tau, \balpha} 
\frac{1}{N} \sum_{i=1}^N (S_i - \tau - \bX_i^\T \balpha)^2 
+ \mu_\init \norm[1]{\balpha},
\] 
$\mu_\init$ and $\mu$ are tuning parameters that can be chosen via the cross-validation or Bayesian information criterion (BIC). {\purple Here, $\balphahat$ is actually an estimator of $\balphaast$, which has the same direction as $\balpha_0$ under the conditions in Proposition \ref{thm:perfect-surrogate}.}

Appending the penalty term \eqref{eq:penalty} to the likelihood 
and replacing $\balpha_0$ with its estimate $\balphahat$, 
we propose to estimate $\bvartheta_0=(\zeta_0,\gamma_0,\bbeta_0\trans)\trans$ by
\[
\wh\bvartheta=(\zetahat, \gammahat, \bbetahat\trans)\trans = 
\argmin_{\bvartheta} 
\frac{1}{n} \sum_{i=1}^n 
\ell(Y_i, \bvartheta \bWvec_i) + 
\lambda_1 \{ \min_\rho \norm[1]{(\bbeta - \rho \balphahat)_{\cAhat}} \} +
\lambda_2 \norm[1]{\bbeta_{\cAhat^c}}, 
\]
where $\cAhat = \supp(\balphahat)$.
The estimators can be equivalently obtained as
\begin{equation}
\label{eq:semisupervised-beta}
\rhohat, \wh\bvartheta = 
\argmin_{\rho,\bvartheta} 
\frac{1}{n} \sum_{i=1}^n
\ell(Y_i, \bvartheta\bWvec_i) + 
\lambda_1 \norm[1]{(\bbeta - \rho \balphahat)_{\cAhat}} +
\lambda_2 \norm[1]{\bbeta_{\cAhat^c}}
\end{equation}
The impact of the tuning parameters $\lambda_1, \lambda_2$ can be understood from
a bias-variance tradeoff viewpoint. When $\lambda_j$'s are large,
$\bbetahat$ tends to be a multiple of $\balphahat$
and thus is an estimator with high bias and low variance.
In contrast, when $\lambda_j$'s are small,
the likelihood term based on the labeled data $\sL$ is
the dominant part,
and hence $\bbetahat$ will have low bias and high variance.
By varying the values of $\lambda_j$'s,
we are able to obtain a continuum connecting these two extremes. In practice, $\lambda_1$ and $\lambda_2$ can be chosen via standard data-driven approaches such as the cross-validation.

\subsection{Computation Details}

The minimization in \eqref{eq:semisupervised-beta} can be solved
with standard software for LASSO estimation.
Let $\bdelta = \bbeta - \rho \balphahat$.
We can re-parametrize the expression above in terms of 
$\rho$, $\zeta$, $\gamma$, and $\bdelta$ as
\[
\zetahat, \gammahat, \rhohat, \bdeltahat = 
\argmin_{\zeta, \gamma, \rho, \bdelta} 
\frac{1}{n} \sum_{i=1}^n 
\ell(Y_i, \zeta + S_i \gamma
+ \rho \bX_i^\T \balphahat + \bX_i^\T \bdelta)
+ \lambda_1 ( \norm[1]{\bdelta_{\cAhat}}
+ \kappa \norm[1]{\bdelta_{\cAhatcomp}} ),
\]
where $\cP = \{1, \dots, p\}$ and $\kappa = \lambda_2 / \lambda_1$. This is a typical LASSO problem with
covariates $(1, S_i, \bX_i^\T \balphahat, \bX_i^\T)^\T$,
parameters $(\zeta, \gamma, \rho, \bdelta)^\T$,
and a weighted $\ell_1$ penalty on the parameters.
Hence it can be solved by 
essentially any algorithm for ALASSO fitting.
In this paper, we use the \texttt{R} package \texttt{glmnet}
\citep{friedman2010}
to compute $\zetahat$, $\gammahat$, $\rhohat$, and $\bdeltahat$,
and construct the final estimator for $\bvartheta_0$ as $\wh\bvartheta = (\zetahat, \gammahat, \bbetahat\trans)\trans$ with $\bbetahat = \bdeltahat + \rhohat \balphahat$.

\section{Theoretical Properties}
\label{sec-theory}

In this section, we present non-asymptotic risk bounds for the PASS estimator. 
We also make theoretical comparisons with the supervised LASSO estimator to shed light on when PASS outperforms the LASSO and where such improvement comes from. 

\subsection{Notations}
A random variable $V$ is sub-Gaussian($\tau^2$)
if $\E\{\exp(\lambda \abs{V})\} \leq 2 \exp(\lambda^2 \tau^2 / 2)$ holds for all $\lambda > 0$.
Throughout, we define
\begin{gather*}
\bU = (\bX^\T, 1)^\T,\;
\bK = \E \big( \bU \bU^\T), \;
\bxi = (\balpha^\T, \tau)^\T,  \bZalpha = (\bX^\T, \bX^\T \balpha, S, 1)^\T, \bG = \E \big( \bZalphaast \bZalphaast^\T \big),\\
\btheta = (\bdelta^\T, \rho, \gamma, \zeta)^\T, \;  \bH = \E \big[
\pi(\bZalphaast^\T \btheta_0) \{1 - \pi(\bZalphaast^\T \btheta_0)\} 
\bZalphaast \bZalphaast^\T \big], 
\end{gather*}
where $\balphaast$ is given by
$(\balphaast{}^\T, \tauast)^\T = \bxiast = \argmin_{\bxi} \E(S - \bU^\T \bxi)^2$, and $\boldsymbol{\Theta}_0=\{\btheta:\bdelta + \rho \balphaast=\bbeta_0,\zeta=\zeta_0,\gamma=\gamma_0\}$. Denote by $\cB_0 = \supp(\bbeta_0)$ and $\cAast = \supp(\balphaast)$. We assume $\|\balpha^*\|_2=1$ without loss of generality since $\balpha^*$ is used to recover only the direction of $\bbeta_0$ in SIM and one can change $\rho$ correspondingly to make any $\bbeta = \bdelta + \rho \balphaast$ invariant to $\|\balpha^*\|_2$. Note that under ($\ref{Logistic}$), any $\btheta_0\in\boldsymbol{\Theta}_0$ minimizes $\E \{\ell(Y, \bZalphaast^\T \btheta)\}$, and due to perfect multicollinearity in $\bZalphaast$,
$\btheta_0$ is not unique. However, any $\btheta_0\in\boldsymbol{\Theta}_0$ corresponds to the unique $\bbeta_0 = \bdelta_0 + \rho_0 \balphaast$ and thus $\bZalphaast^\T \btheta_0 = 
\zeta_0 + S \gamma_0 + \bX^\T \bbeta_0 = \bvartheta_0\trans\bWvec$ is well-defined. Moreover, any quantity depending on $\btheta_0$
through $\bZalphaast^\T \btheta_0$ is well-defined. Since the main results in this section depend on $\btheta_0$ solely through $\bZalphaast^\T \btheta_0$, we will use $\btheta_0$ to represent any $\btheta\in\boldsymbol{\Theta}_0$ for simplicity.

For $\btheta = (\bdelta^\T, \rho, \gamma, \zeta)^\T$, define 
$
\Omega(\btheta) = 
\lambda_0 (\abs{\rho} + 
\abs{\gamma} + \abs{\zeta}) +
\lambda_1 \norm[1]{\bdelta_\cAast} + 
\lambda_2 \norm[1]{\bdelta_\cAastcomp}
$, $\Deltaalpha=2\mu_\init q^*/\varphi^2$ and 
$\Pi(\btheta) = \abs{\rho}$,
where $\lambda_0 = 36 B \{ \log(6 / \epsilon) / n \}^{1/2}$.
To introduce the oracle $\bthetaast$, we define the oracle risk function as:
\begin{equation}
\label{eq:excess-risk}
\begin{split}
\sE(\btheta, \cSplus, \cSminus) = 
& \E \ell(Y, \bZalphaast^\T \btheta) - 
\E \ell(Y, \bZalphaast^\T \btheta_0) \\ 
& + 256 \frac{\kappa(\cSplus)^2 \abs{\cSplus}}{\varpi \psi(\cSplus)} +
8 \lambda_1 \norm[1]{\btheta_{\cSminus \cap \cAast}} + 
8 \lambda_2 \norm[1]{\btheta_{\cSminus \cap (\cAastcomp)}} +
8 \lambda_1 \Deltaalpha \Pi(\btheta),
\end{split}
\end{equation}
where
\begin{align*}
\psi(\cSplus) &= 
\inf_{\bv: \Omega(\bv_{\cSminus}) \leq 3 \Omega(\bv_\cSplus)} 
\frac{\bv^\T \bG \bv}{\bv_\cSplus^\T \bv_\cSplus}, \\
\kappa(\cSplus) &= \begin{cases}
\lambda_0, \quad \text{if } 
\cSplus \cap \cAast = \emptyset \text{ and } 
\cSplus \cap (\cAastcomp) = \emptyset \\
\lambda_2, \quad \text{if } 
\cSplus \cap \cAast = \emptyset \text{ and } 
\cSplus \cap (\cAastcomp) \neq \emptyset \\
+\infty, \quad \text{if } 
\cSplus \cap \cAast \neq \emptyset 
\end{cases}
\end{align*}
Define 
$ \bthetaast = (\bdeltaast{}^\T, \rhoast, \gammaast, \zetaast)^\T$, 
$\cSplusast$ and $\cSminusast$ as the solution to
\begin{align*}
&\argmin_{\{\btheta, \cSplus, \cSminus\}: \cSplus \cap \cSminus = \emptyset,\;
\cSplus \cup \cSminus = \supp(\btheta)\cup\cPbar,\;
\cSplus \supseteq \cPbar,
\text{ and }
\norm[2]{\bG^{1/2} (\btheta - \btheta_0)} \leq \sigma,
} 
\sE(\btheta, \cSplus, \cSminus) 
\end{align*}
where $\cPbar = \{p + 1, p + 2, p + 3\}$.
Let $\cSast = \cSplusast \cup \cSminusast = \supp(\bthetaast)\cup\cPbar$, $\kappaast = \kappa(\cSplusast)$, 
and $\bbetaast = \bdeltaast + \rhoast \balphaast$. Intuitively, one may view $\cSplus$ as the union set of unpenalized predictors and 
the predictors with large coefficients but not recovered by $\cAast$. While $\cSminus$ can be viewed as the union set of predictors
with small nonzero coefficients and the predictors recovered by $\cAast$.
Partitioning the support of $\btheta$ into $\cSplus$ and $\cSminus$ 
is inspired by \citet[][Section~6.2.4]{buhlmann2011},
which leads to a refined bound.

{\purple

\subsection{Main result}

We first establish the risk bounds for the PASS estimator in the following theorem. Its proof can be found in Section S2 of the Supplementary Materials.

\setcounter{theorem}{0}
\begin{theorem}
For any $\epsilon > 0$, 
if the assumptions (A1)--(A8) (introduced in Section S2.1 of the Supplementary Materials) hold,
the following inequalities hold simultaneously with probability at least 
$1 - 10 \epsilon$:
\begin{alignat*}{2}
& \mbox{Excess risk:} \quad &\E \ell(Y, \bZalphahat^\T \bthetahat) -  
\E \ell(Y, \bZalphaast^\T \btheta_0) 
&\leq \Xi, \\
& \mbox{Linear prediction error:} \quad &\E (\bZalphahat^\T \bthetahat - \bZalphaast^\T \btheta_0)^2 
&\leq \Xi / \varpi, \\
& \mbox{Probability prediction error:} \quad &\E \{ \pi(\bZalphahat^\T \bthetahat) - \pi(\bZalphaast^\T \btheta_0) \}^2 
&\leq \Xi / \varpi,
\end{alignat*}
where $\varpi$ is a positive constant defined in (A2), $\Xi = 64 \sE(\bthetaast, \cSplusast, \cSminusast)$, and $\sE$ is an oracle risk function as defined in equation \eqref{eq:excess-risk}.
\label{thm:1}
\end{theorem}

\begin{remark}
As detailed in Section S2.1 of the Supplement Material, Assumptions (A1)--(A8) are imposed on tail behaviour of the regression residuals, regularity of the design matrix, minimum signal strength of $\balphaast$, sample sizes and rates of the tuning parameters. These assumptions are commonly used conditions in the theoretical literature of LASSO, such as the sub-Gaussian variable condition and the restricted eigenvalue condition; see e.g. \cite{vandegeer2009,bickel2009,buhlmann2011}. 
\end{remark}

\begin{remark}\label{remark-lastterm}
The last term of the risk bound $\sE(\bthetaast, \cSplusast, \cSminusast)$ is of order $O(\lambda_1 \Deltaalpha \abs{\rhoast})$, which reflects the estimation error in $\balphahat$. Following Lemma S8 in the Supplement, one can show that $\Deltaalpha=O_p(N^{-1/2}|\cAast|)$. All the other terms in $\Xi$ describe the estimation error in $\bthetahat$ as if $\balphahat$ is replaced with $\balphaast$. When $N$ is sufficiently large, $O(\lambda_1 \Deltaalpha \abs{\rhoast})$ is typically negligible relative to other terms. Specifically, if $N\gg n|\cAast|^2\log (p)$, $O(\lambda_1 \Deltaalpha \abs{\rhoast})=O(\{Nn\}^{-1/2} \log(p)^{1/2}|\cAast|)=o(n^{-1})$. In general, as long as $N \gg \max(n,p)$ and $\balphaast$ is not much denser than $\bbeta_0$ as in the typical EHR application cases, $O(\lambda_1 \Deltaalpha \abs{\rhoast})$ is dominated by the risk of the supervised LASSO estimator and even the supervised oracle estimator obtained under the knowledge of $\supp(\bbeta_0)$. 
\end{remark}

To gain a better understanding of how the key quantity $\Xi$
in Theorem~\ref{thm:1} changes with respect to the similarity between the prior information $\balphaast$ and the target $\bbeta_0$, we shall discuss several specific cases in Section \ref{sec:thm:case}, based on the risk bound derived in Theorem~\ref{thm:1}.
}

{\purple
\subsection{Specific Cases}\label{sec:thm:case}
Following Remark \ref{remark-lastterm}, we focus our discussions on the settings where $N$ is sufficiently large such that the last term of the risk bound is negligible. We consider three different scenarios as illustrated in Figure \ref{fig:thm}:  (Case 1) $\balpha^*$ recovers both the support and direction of $\bbeta_0$;  (Case 2) $\balpha^*$ almost recovers the support of $\bbeta_0$ but has a substantially different direction from $\bbeta_0$;  (Case 3) $\balpha^*$ fails to recover the support of $\bbeta_0$ (let alone its direction) and provides poor information. These three cases depict perfect, good, and poor qualities of the prior information $\balpha^*$ in recovering the support and direction of  $\bbeta_0$. Next, we rigorously characterize the three cases by properly specifying the parameters $\rho$, $\bdelta$, $\cSplus$, and $\cSminus$, and derive the convergence rate of $\Xi$, the risk bound of the PASS estimator, based on Theorem \ref{thm:1}.

\begin{case}
Let $\rhobar = \min_\rho \norm[1]{\bbeta_0 - \rho \balphaast}$, 
$\bdeltabar = \bbeta_0 - \rhobar \balphaast$,
$\bthetabar = (\bdeltabar^\T, \rhobar, \gamma_0, \zeta_0)^\T$,
$\cSplusbar = \cPbar$
and $\cSminusbar = \supp(\bdelta_0)$. If $\balpha^*$ successfully recover the support and direction of $\bbeta_0$ (see the left panel in Figure \ref{fig:thm}), $\cSminusbar\approx\emptyset$ and $\norm[1]{\bdelta_0} \approx 0$. Since $\norm[2]{\bG^{1/2} (\wb{\btheta} - \btheta_0)}=0$ and $\cSplusbar\cap\cAast=\emptyset$, we have
$
\Xi = O\{ \sE(\bthetabar, \cSplusbar, \cSminusbar) \}
$ by the definition of $\bthetaast$.
Hence by Theorem \ref{thm:1}, the excess risk of $\bthetahat$
\[
\Xi = O_p(\lambda_0^2 + 
\lambda_1 \norm[1]{\bdeltabar_{\cSminusbar \cap \cAast}} +
\lambda_2 \norm[1]{\bdeltabar_{\cSminusbar \cap \cAast{}^c}})\approx O_p(\lambda_0^2) =O_p(\ninv),
\]
recalling that $\lambda_0 = O(\nnhalf)$.
\label{case:1}
\end{case}
As a standard result \citep{negahban2009unified}, the rate of the excess risk of the supervised LASSO estimator is either $O_p\{n^{-1}\log (p)|\cB_0|\}$ or $O\{\nnhalf \log(p)^{1/2}\norm[1]{\bbeta_0}\}$. These two rate bounds are established under different sparsity norms of $\bbeta_0$, and generally comparable, e.g. when order of average magnitude of the non-zero entries in $\bbeta_0$ is $\nnhalf \log(p)^{1/2}$. In comparison with them, $O_p(\ninv)$, the risk rate of PASS in Case \ref{case:1}, is much more refined. Further, $O_p(\ninv)$ is actually the rate of the estimator of a low (fixed) dimensional logistic regression. Thus, if $\bbeta_0$ is very close to a multiple of $\balphaast$, PASS could outperform the vanilla LASSO and be comparable with a low dimensional regression in terms of the convergence rate. This big gain is owing to the use of $N$ unlabeled data to obtain the direction of $\bbeta_0$, and thus reduce the high dimensional regression to a low dimensional one where only the intercept and the scalar of $\bbeta_0$ need to be estimated.

\begin{case}
Consider the same choice of $\bthetabar$, $\cSplusbar$ and $\cSminusbar$ as in Case \ref{case:1}. If $\balpha^*$ recovers the support but not the direction of $\bbeta_0$ (see the middle of Figure \ref{fig:thm}), we will only have $\norm[1]{\bdeltabar_{\cSminusbar \cap \cAast{}^c}} \approx 0$ but not $\norm[1]{\bdelta_0} \approx 0$. Then by Theorem \ref{thm:1}, the excess risk of PASS is
\[
\Xi= O_p(\lambda_0^2 + 
\lambda_1 \norm[1]{\bdeltabar_{\cSminusbar \cap \cAast}} +
\lambda_2 \norm[1]{\bdeltabar_{\cSminusbar \cap \cAast{}^c}})\approx O_p\{\nnhalf \log(q^\ast)^{1/2}\norm[1]{\bdeltabar_{\cSminusbar \cap \cAast}}\},
\]
recalling that $\lambda_1 = O\{\nnhalf \log(q^\ast)^{1/2}\}$. 
\label{case:2}
\end{case}
In Case \ref{case:2}, the convergence rate of the excess risk of PASS is still better than that of the supervised LASSO estimator when $q^*\ll p$:
\[
O\{\nnhalf \log(q^*)^{1/2}\norm[1]{\bdeltabar_{\cSminusbar \cap \cAast}}\} \ll O\{\nnhalf \log(p)^{1/2}\norm[1]{\bbeta_0}\},
\]
by $\norm[1]{\bdeltabar_{\cSminusbar \cap \cAast}}\leq \min_\rho \norm[1]{\bbeta_0 - \rho \balphaast}\leq\norm[1]{\bbeta_0}$. Namely, if $\balphaast$ might not recover the direction of $\bbeta_0$ very well but the prior information $\cAast=\supp(\balphaast)$ is sparse and covers $\supp(\bbeta_0)$ successfully, which is reflected as $\cSplusbar = \cPbar$, the PASS estimator still benefits from the prior information. This is because recovering the support of $\bbeta_0$ reduces the dimensionality of the empirical errors needed to be controlled from $p$ to $q^*=|\cAast|$. In this case, it is also interesting to compare the proposed PASS estimator with the prior LASSO (pLASSO) procedure of \citet{jiang2016}. When $\supp(\balphaast)$ and $\supp(\bbeta_0)$ are close but the directions of $\balphaast$ and $\bbeta_0$ are quite different, the pLASSO procedure is unable to utilize this information and will only result in the same convergence rate as supervised LASSO, as shown to be essentially slower than that of PASS.




\begin{case}
Let $\rhobar = 0$, 
$\bdeltabar = \bbeta_0$,
$\bthetabar = (\bdeltabar^\T, \rhobar, \gamma_0, \zeta_0)^\T$, $\cSplusbar = \cPbar\cup(\cB_0\setminus\cAast)$ and $\cSminusbar=\cB_0\setminus\cSplusbar$. If $\balpha^*$ fails to recover the support of $\bbeta_0$, i.e.  $\cAast\cap \cB_0\approx\emptyset$ and $\norm[1]{\bbeta_{0,\cAast\cap\cB_0}}\approx 0$, we have $\norm[1]{\bdeltabar_{\cSminusbar}}\leq\norm[1]{\bdeltabar_{\cAast}}\approx\norm[1]{\bbeta_{0,\cAast\cap\cB_0}}\approx 0$. Then again using Theorem \ref{thm:1}, 
\[
\Xi = O(\lambda_2^2 \abs{\cSplusbar}+ 
\lambda_1 \norm[1]{\bdeltabar_{\cSminusbar \cap \cAast}} +
\lambda_2 \norm[1]{\bdeltabar_{\cSminusbar \cap \cAast{}^c}})\approx O_p\{\ninv \log(p)|\cB_0|\},
\]
recalling that $\lambda_2 = O\{\nnhalf \log(p)^{1/2}\}$.
\label{case:3}
\end{case}

In Case~\ref{case:3},
the excess risk of the PASS is of the same order as
that of supervised LASSO. Therefore the PASS approach is robust against 
low-quality prior information that recovers neither the direction nor the support of $\bbeta_0$. This benefit is a result of using a data-adaptive parameter $\rho$ to control the influence of the prior information on the estimator.

\begin{figure}[htb!]
\centering
\includegraphics[width=0.95\textwidth]{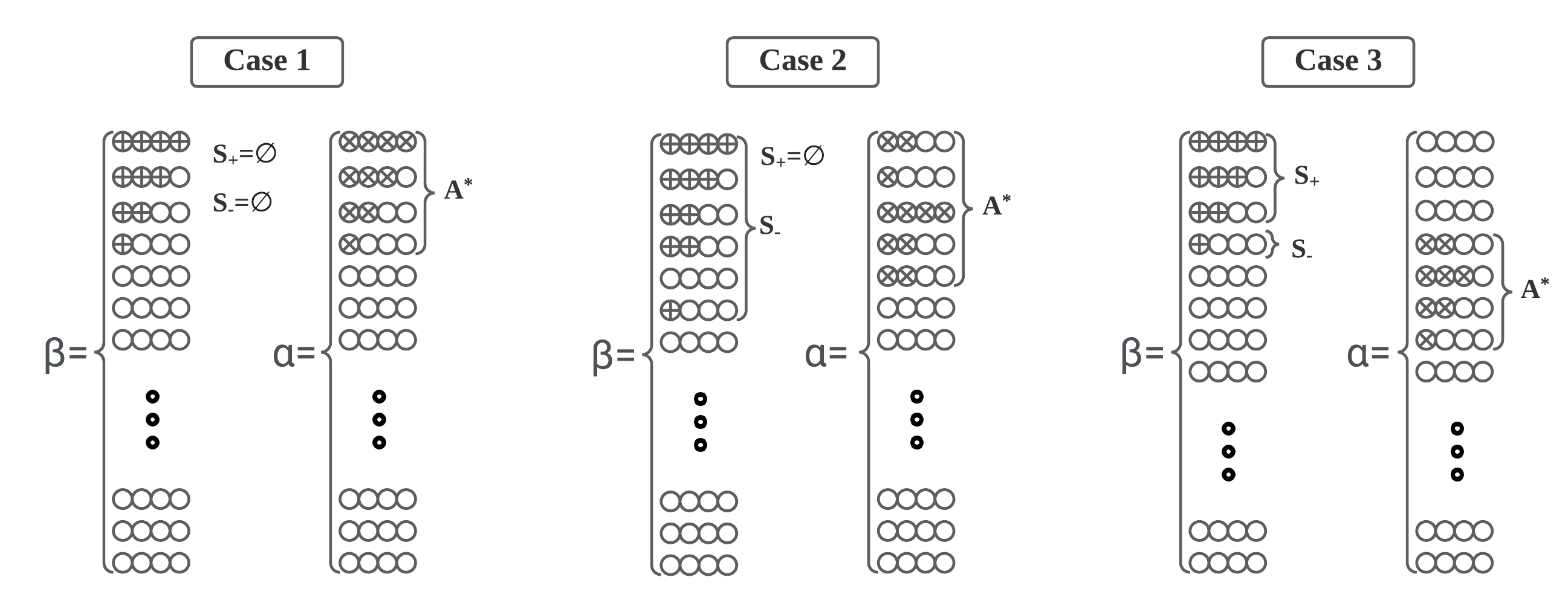}
\caption{{\purple \label{fig:thm} Sketch diagrams of the coefficients $\bbeta_0$ and $\balpha^*$ in the three cases of Section \ref{sec:thm:case}. Labels S$_{\sf \scriptscriptstyle +}$, S$_{\sf \scriptscriptstyle -}$, and A$^*$ in the diagrams represent $\cSplusbar\setminus\cPbar$, $\cSminus$, and $\cAast$ as chosen and defined in Cases \ref{case:1}--\ref{case:3}. $\bbeta_0$ and $\balpha^*$ are aligned for comparison of their directions and supports. Symbols $\bigcirc$, $\vdots$, $\bigoplus$, and $\bigotimes$ represent zero, ellipsis of zero coefficients, a unit of magnitude in $\bbeta_0$, and a unit of magnitude in $\balpha^*$ respectively. Case \ref{case:1} (presented in the left panel): $\balpha^*$ recovers both the support and direction of $\bbeta_0$. Case 2 (middle): $\balpha^*$ (nearly) recovers the support but not the direction $\bbeta_0$. Case 3 (right): $\balpha^*$ fails to recover the support of $\bbeta_0$.}}
\end{figure}


}

\section{Simulation Studies}\label{sec:sim}

\subsection{Main setups}\label{sec:sim:main}

{\purple 
We conducted extensive simulation studies to examine the finite-sample performance of the PASS estimator and compare to existing estimators. We first consider the case where the logistic model for $Y \mid S, \bX$ is correctly specified, $S \mid \bX$ follows an SIM, and $\bX$ is near elliptical, but  the similarity between $\balpha_0$ and $\bbeta_0$ varies. 
Since EHR features are often zero inflated and skewed count variables, we generated $\bX_{500\times 1}$ from
$$\bX_i = h(\bZ_i), \quad \bZ_i\sim N(\zero, \bSigma_{\bZ}), \quad h(t) = \log(1 + [e^t]),$$ 
where $[u]$ denotes the integer nearest to $u$, $\bSigma_{\bZ} = (\sigma_{i,j})_{i, j = 1}^p$
and $\sigma_{i,j} = 4(0.5)^{\abs{i-j}}$. Here $[e^{Z_{ij}}]$ mimics the skewed raw EHR feature, which is typically transformed via
$t \to \log(1 + t)$ prior to model fitting.
We then generated the surrogate $S$ from a SIM of $\bX$:
$$S_i = h(1 + \bX_i^\T \balpha_0 + \epsilon_i), \quad \mbox{with}\quad \epsilon_i \sim N(0, 2^2).$$ 
Following the model assumption ($\ModelY$), the disease status $Y_i$ was generated from  
$$
\logit \Pr(Y_i = 1 \mid \bW_i) = -4 + 0.5 S_i + \bX_i^\T \bbeta_0 .
$$ 
To mimic different qualities of the prior information one could encounter in practice, we design six scenarios with different similarities between the true $\bbeta_0$ and $\balpha_0$: 
\begin{alignat*}{3}
& \text{\RomanNum{1}}: &\quad&
 \balpha_0 = (\ba_1^\T, \ba_2^\T, \zero_{p-10}^\T)^\T,
&\quad& \bbeta_0 = 1.5 (\ba_1^\T, \ba_2^\T, \zero_{p-10}^\T)^\T; \\
& \text{\RomanNum{2}}: &\quad&
\balpha_0 = (\ba_1^\T, \ba_2^\T, \zero_{p-10}^\T)^\T, 
&\quad& \bbeta_0 = 1.5 (\ba_1^\T + \bd_1^\T, \ba_2^\T + \bd_2^\T, \zero_{p-10}^\T)^\T;\\
& \text{\RomanNum{3}}: &\quad&
\balpha_0 = (\ba_1^\T, \ba_2^\T, \ba_2^\T, \ba_2^\T, \zero_{p-20}^\T)^\T, 
&\quad&  \bbeta_0 = 1.5 (\ba_1^\T + \bd_1^\T, \ba_2^\T + \bd_2^\T, \zero_{p-10}^\T)^\T; \\
& \text{\RomanNum{4}}: &\quad&
\balpha_0 = (\ba_1^\T, \zero_{p-5}^\T)^\T, 
&\quad& \bbeta_0 = 1.5 (\ba_1^\T + \bd_1^\T, \ba_2^\T + \bd_2^\T, \zero_{p-10}^\T)^\T;\\ 
& \text{\RomanNum{5}}: &\quad&
\balpha_0 = (\ba_1^\T, \ba_2^\T, \zero_{p-10}^\T)^\T, 
&\quad& \bbeta_0 = 1.5 (\ba_2^\T, \ba_1^\T, \zero_{p-10}^\T)^\T; \\
& \text{\RomanNum{6}}: &\quad&
\balpha_0 = (\ba_1^\T, \ba_2^\T, \zero_{p-10}^\T)^\T, 
&\quad& \bbeta_0 = 1.5 (\ba_2^\T, \zero_5, \ba_1^\T, \zero_{p-15}^\T)^\T.
\end{alignat*}
where
\begin{align*}
\ba_1 &= (0.5, 1, -0.8, 0.6, 0.2)^\T, &
\bd_1 &= (-0.05, -0.5, 1.4, 0.5, -0.6)^\T, \\
\ba_2 &= (0.1, -0.2, -0.2, 0.2, 0.7)^\T, &
\bd_2 &= (0.02, 0.05, 0.02, -0.02, -0.05)^\T.
\end{align*}
Our specifications of $\bbeta_0$ and $\balpha_0$ are motivated by the three key specific cases introduced in Section \ref{sec:thm:case} and illustrated in Figure \ref{fig:thm}.} 
Scenario~\RomanNum{1} is the ideal case where
$\bbeta_0$ and $\balpha_0$ have identical direction.
In Scenario~\RomanNum{2}, most of the components of $\bbeta_0$ 
differ slightly from a scalar multiple of $\balpha_0$, 
while a few components differs substantially.
Scenarios~\RomanNum{1} and~\RomanNum{2} are designed to examine
the performance of PASS estimator when the prior information is 
highly or somewhat reliable.
In Scenario~\RomanNum{3}, $\balpha_0$ is denser than $\bbeta_0$
and contains quite a few weak signals. 
On the contrary, in Scenario~\RomanNum{4}
$\bbeta_0$ is denser than $\balpha_0$.
In Scenario~\RomanNum{5}, the magnitude of $\balpha_0$
and $\bbeta_0$ are quite different, whereas they still share
the same support.
Scenarios~\RomanNum{3}, \RomanNum{4} and~\RomanNum{5} 
are designed to examine the performance of PASS estimator
with respect to different degree of accuracy of the support information.
In Scenario~\RomanNum{6}, both the magnitude and the support of 
$\balpha_0$ and $\bbeta_0$ differs substantially,
which means the unlabeled data provides little information.
This scenario allows us to see whether the PASS estimator
is robust against unreliable prior information. {\purple See Figure S1 in the Supplementary Materials for a visualization of $\bbeta_0$ and $\rho\balpha_0$ across different scenarios.} 

We compare PASS to following existing methods: (1) supervised LASSO penalized logistic regression with $n$ training samples (LASSO$_n$); (2) supervised ALASSO penalized logistic regression with $n$ training samples, denoted by ALASSO$_n$; (3) the \SSprior{}  estimator  as described in section \ref{sec-model}; and (4) two variants of pLASSO estimators as proposed in \citet{jiang2016}. For pLASSO, we  fit a penalized logistic model with an LASSO penalty imposed on predictors outside $\supp(\balphahat)$, as in equation~(8) of \citet{jiang2016}, and then use the predicted probability from that model as $Y_i^\text{p}$ 
in equation~(7) of \citet{jiang2016}, denoted by pLASSO$^1$; (2) use the predicted probability given by the \SSprior{} approach 
as $Y_i^\text{p}$ in equation~(7) of their paper, denoted by pLASSO$^2$.

{Throughout, we let $N = 10000$ and let $\nu = 1$ in the ALASSO weights. We use Bayesian information criterion (BIC) to select $\mu_\init$ and $\mu$ in the estimation of $\balpha$ due to large $N$, and use $10$-fold cross-validation to select  
$\lambda_1$, $\lambda_2$ for the estimation of $\bbeta$, so that the phenotype model is tuned towards prediction performance. We quantify the average prediction performance of the estimated linear score, $\widetilde{\bvartheta}\trans\bWvec$, with $\widetilde{\bvartheta}$ obtained via different methods in an independent test data with size $10000$. For each choice of $\widetilde{\bvartheta}\trans\bWvec$, we consider the area under the receiver operating characteristic curve (AUC) for classifying $Y$, the excess risk (ER) as defined in Section~\ref{sec-theory}, and the mean squared error of the predicted probabilities (MSE-P)
which is the mean squared differences between the predicted probability and the true probability. We  summarize results based on 1000 simulated datasets for each configuration. 
}

\begin{figure}[htbp]
\centering
\includegraphics[width=0.325\textwidth]{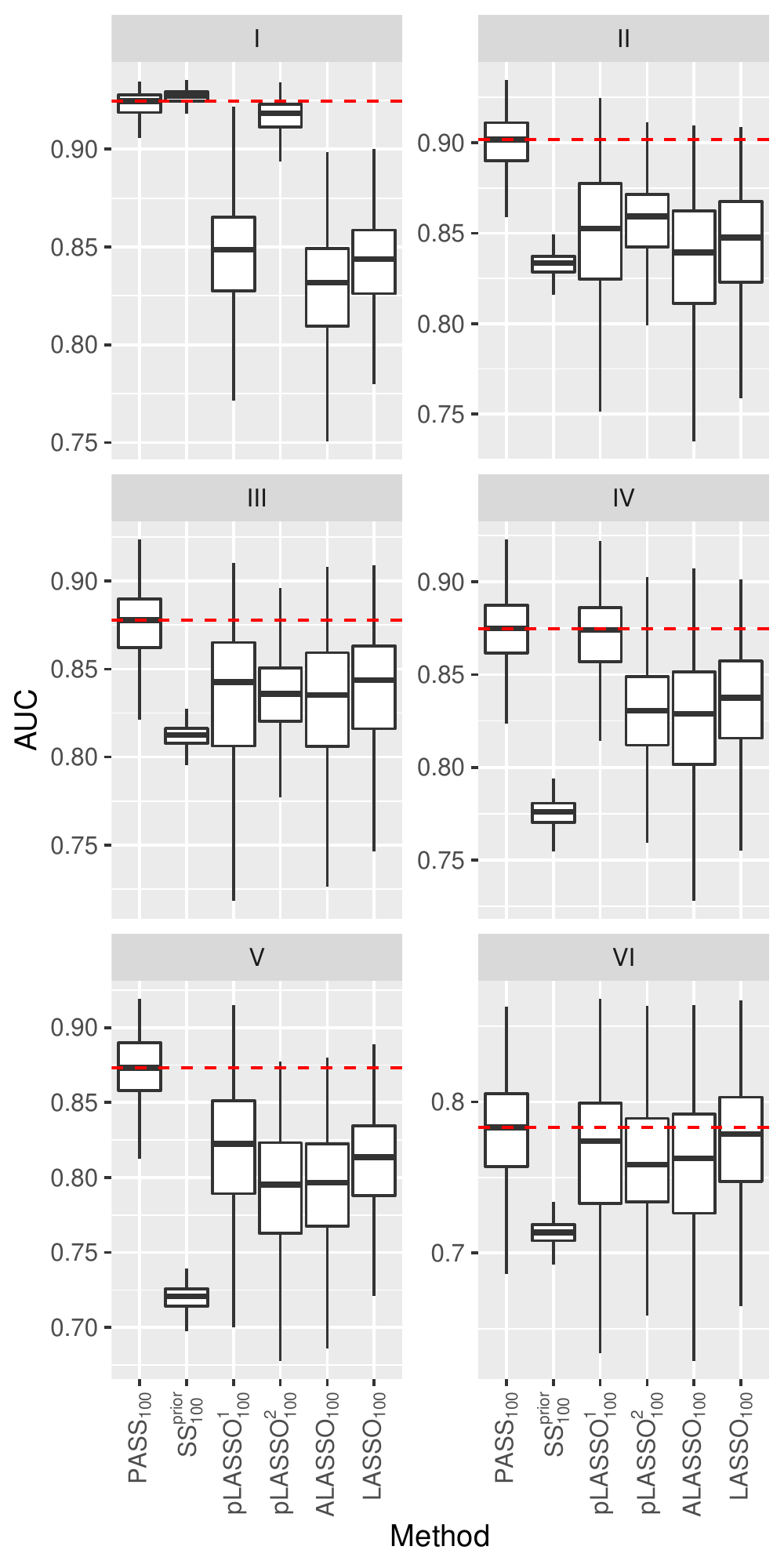}
\includegraphics[width=0.325\textwidth]{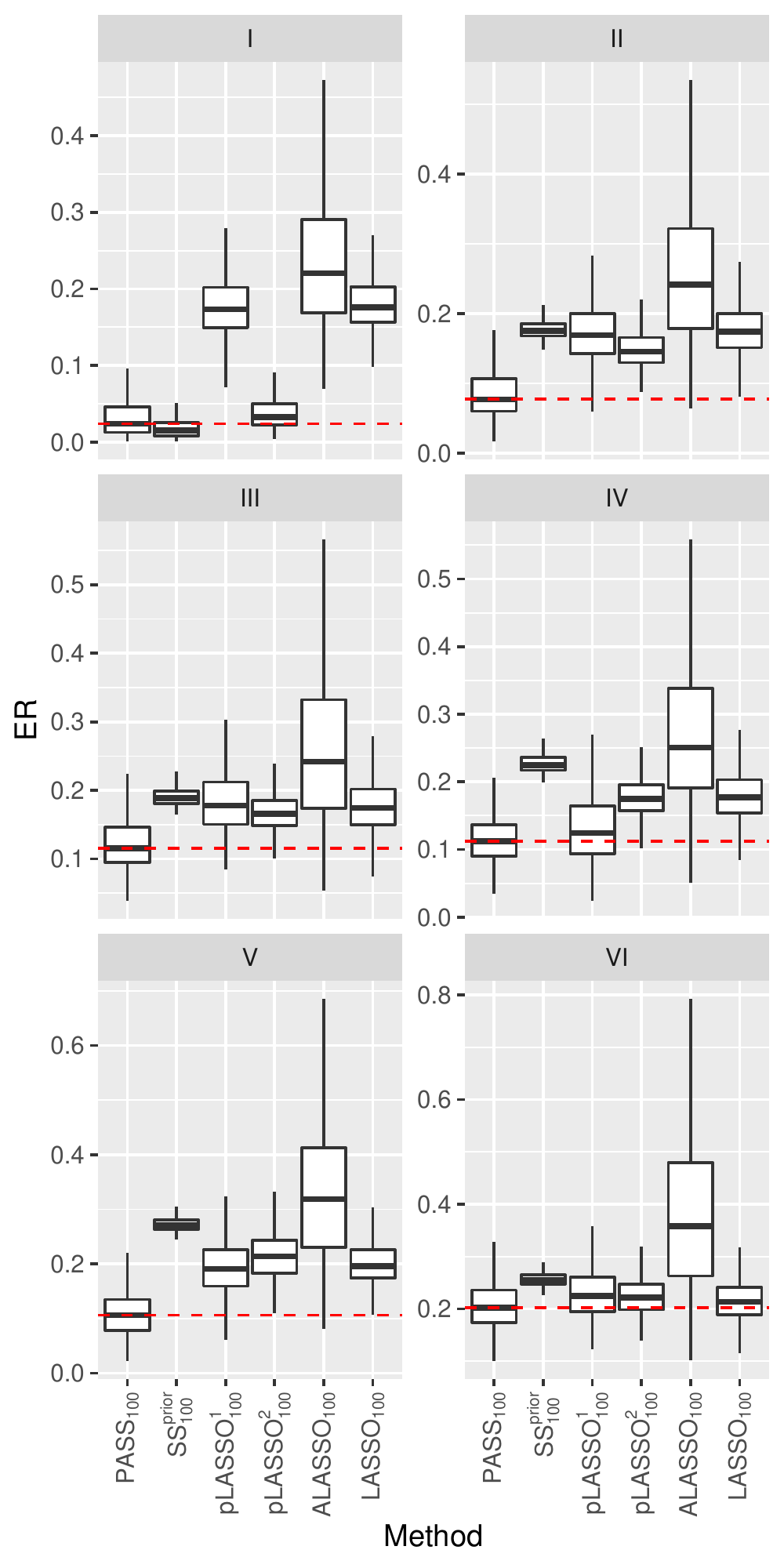}
\includegraphics[width=0.325\textwidth]{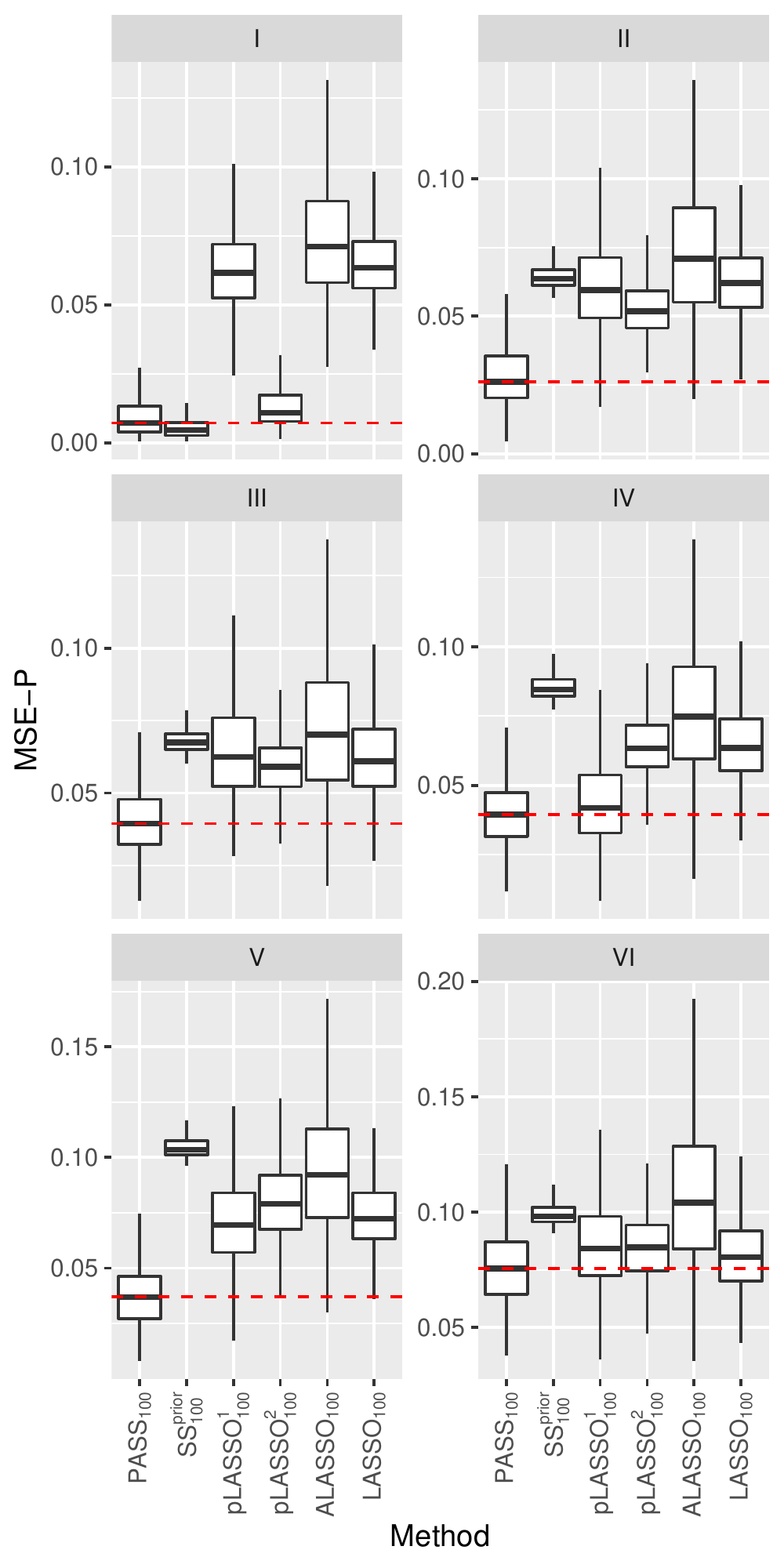}
\caption{AUC (left), ER (middle) and MSE-P (right) evaluated on the test set for simulation studies under Scenarios I--VI.
Outliers are not drawn. {\purple Mean performance of the PASS approach are marked using dashed lines for ease of comparison.} The size of the labeled data is fixed at $n = 100$.}
\label{fig:r1}
\end{figure}

In Figures~\ref{fig:r1}, we compare prediction measures for estimators obtained with $n = 100$. 
In Scenario~\RomanNum{1} where the directions of $\bbeta_0$
and $\balpha_0$ coincide, the \SSprior{} approach 
performs the best as expected, yet the proposed PASS method attained very 
similar accuracy followed by pLASSO$^2$ which performed only slightly worse. 
When the directions of $\bbeta_0$ and $\balpha_0$ are
somewhat different as in Scenario~\RomanNum{2}, the \SSprior{} and the pLASSO estimators 
deteriorate quickly. In contrast, the PASS estimator maintains high accuracy and outperforms all 
competing estimators substantially. We observe qualitatively similar patterns for 
Scenarios~\RomanNum{3} and \RomanNum{4} under which $\balpha_0$ 
and $\bbeta_0$ have somewhat different support. 
No matter whether $\balpha_0$ is denser than $\bbeta_0$ 
as in Scenario~\RomanNum{4}, or $\bbeta_0$ is denser than
$\balpha_0$ as in Scenario~\RomanNum{5},
the PASS method consistently outperforms the supervised estimators.
Additionally, the performances of the \SSprior{}
and pLASSO approaches are not quite satisfactory.
In Scenario~\RomanNum{5}, $\bbeta_0$ and $\balpha_0$ 
have the same support but are quite different in terms of magnitude.
The proposed method managed to utilize the same-support information,
whereas the pLASSO approaches failed to do so.
Finally, the goal of Scenario~\RomanNum{6} is to examine the 
robustness of the methods when $\bbeta_0$ and $\balpha_0$
differs a lot, possibly due to the use of an inappropriate surrogate.
The PASS estimator performs similarly to the supervised estimators, indicating that 
our procedure is indeed adaptive to how well the data support the prior assumption.
Across all scenarios, the ALASSO approach performs slightly worse than LASSO,
possibly due to the presence of some small nonzero coefficients
in $\bbeta_0$.


\begin{figure}[htbp]
\centering
\includegraphics[width=0.325\textwidth]{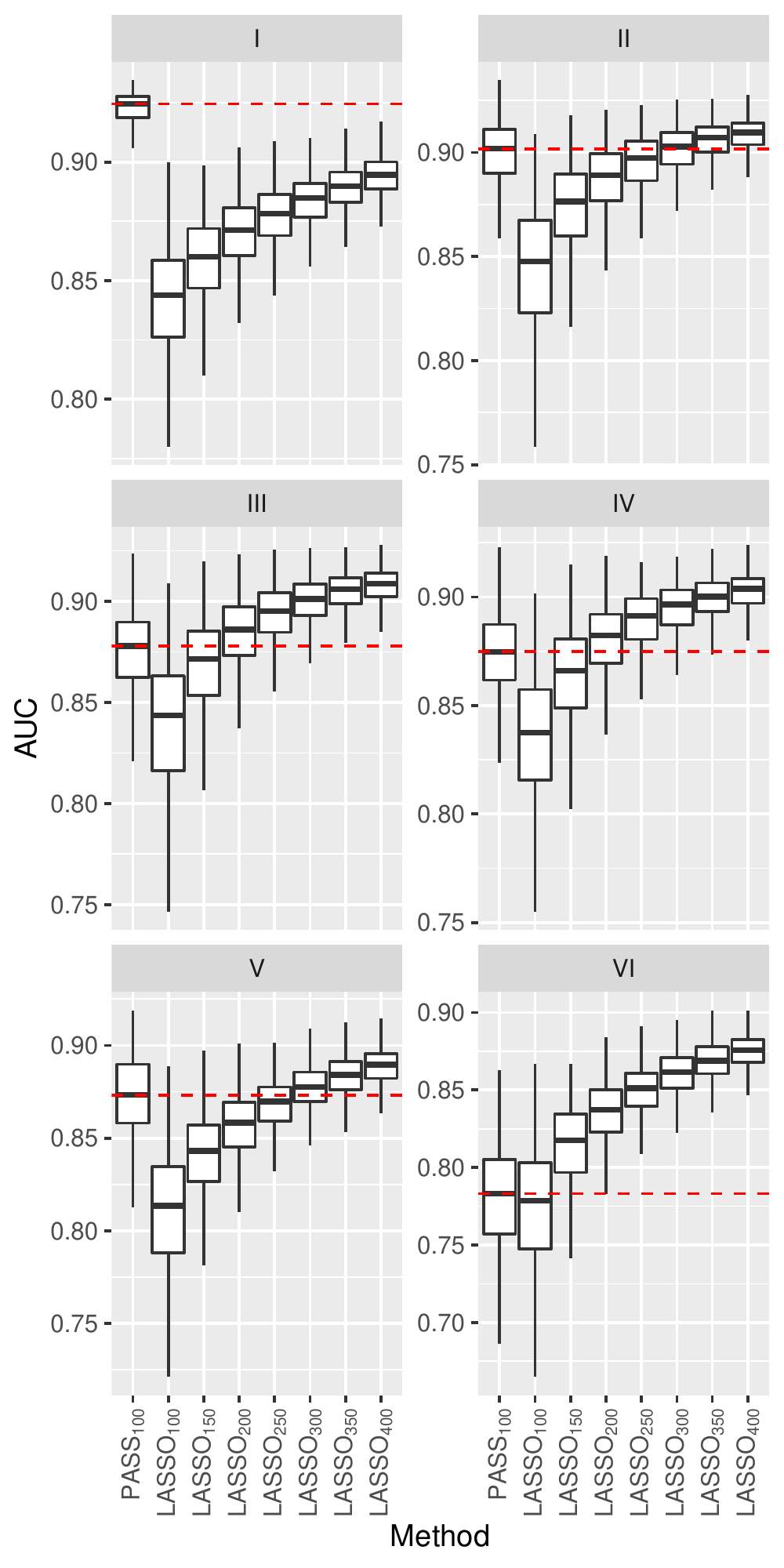}
\includegraphics[width=0.325\textwidth]{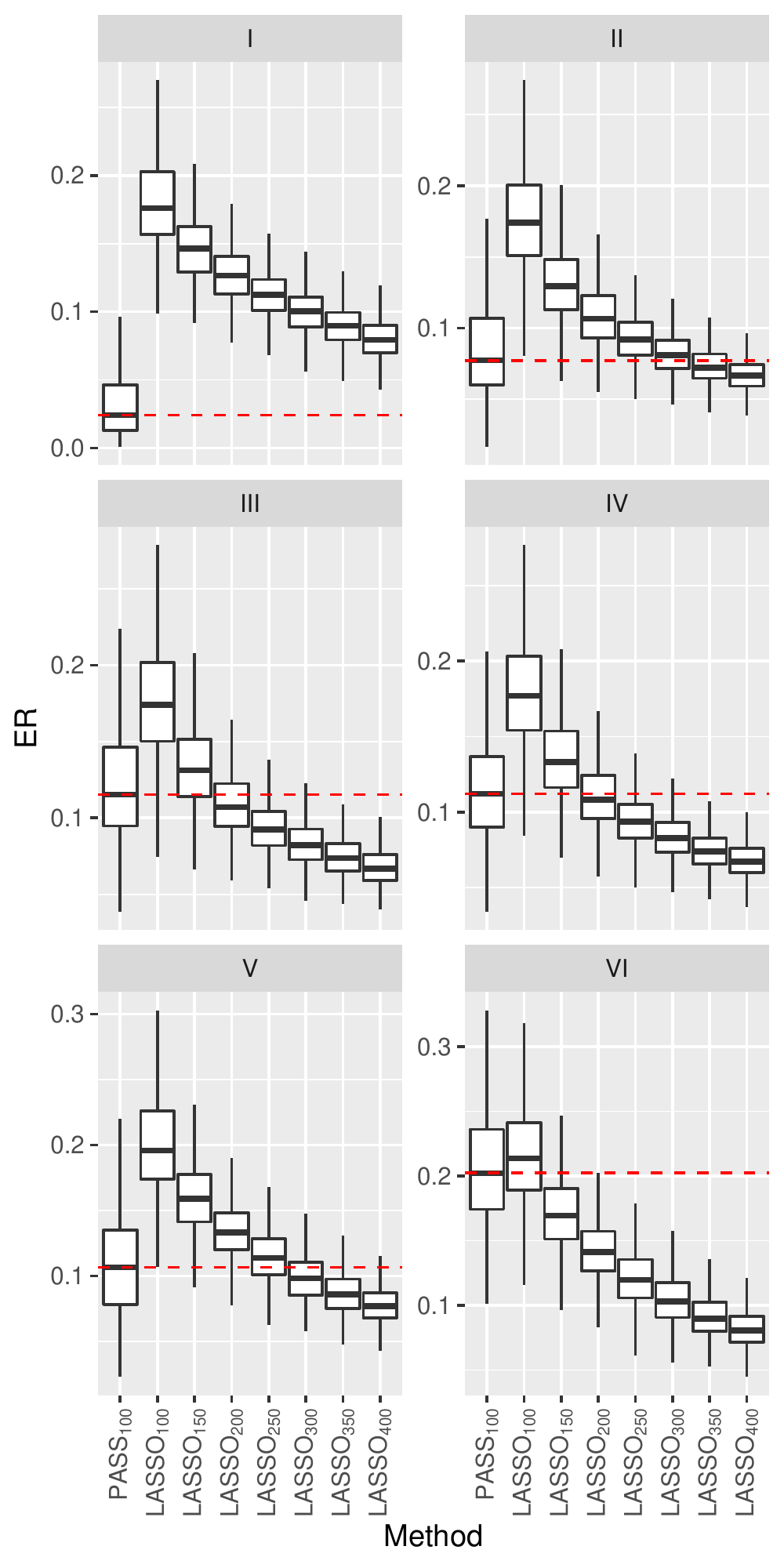}
\includegraphics[width=0.325\textwidth]{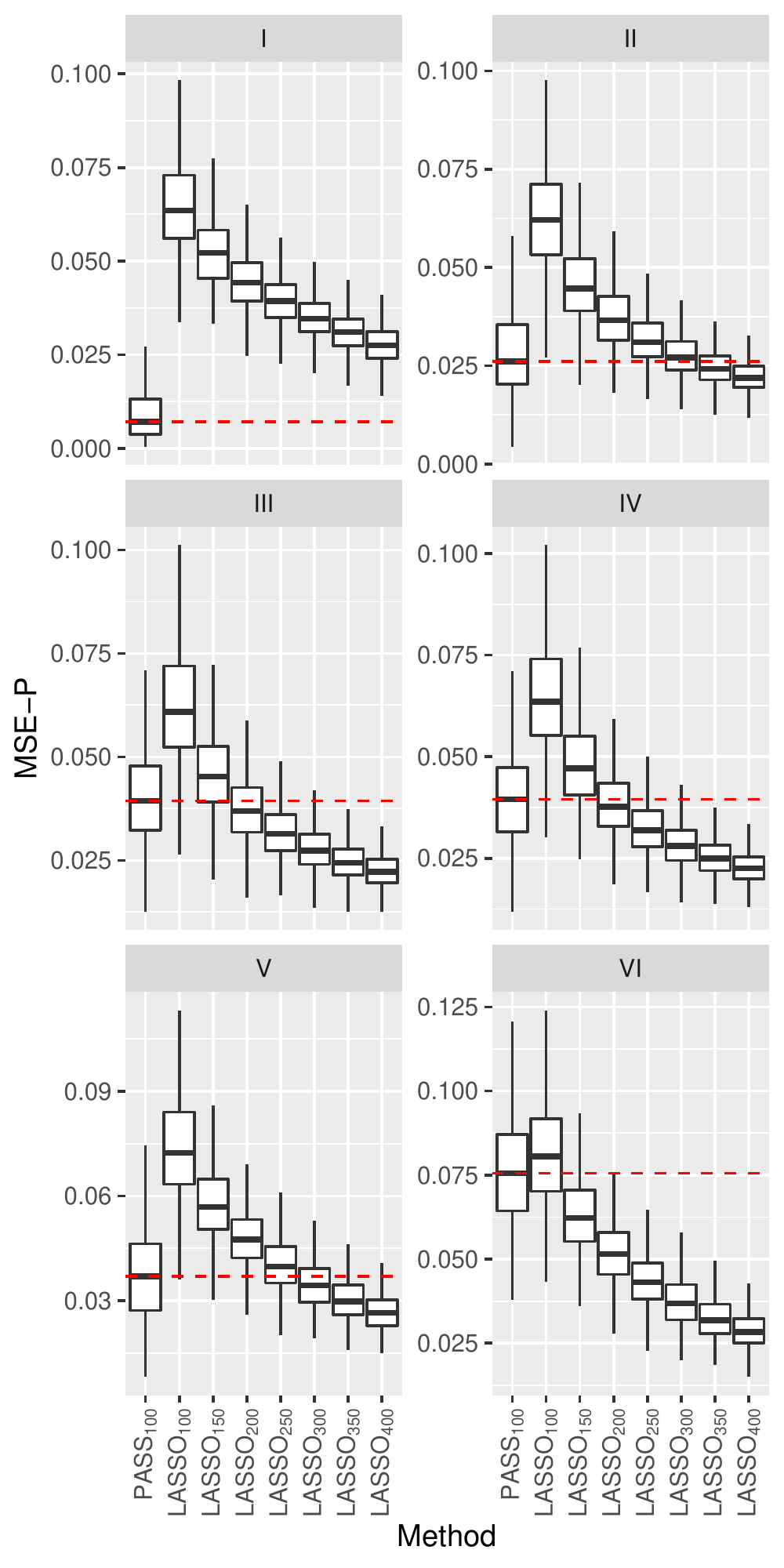}
\caption{AUC (left), ER (middle) and MSE-P (right) evaluated on the test set for simulation studies under Scenarios I--VI.
Outliers are not drawn. {\purple Mean performance of the PASS approach are marked using red dash lines for ease of comparison.} The size of the labeled data is $n = 100$ for PASS, while it varies for LASSO, 
as indicated in the subscripts.}
\label{fig:r2}
\end{figure}

In Figures~\ref{fig:r2}, we present the AUC, ER and MSE-P of the PASS estimator trained with $n = 100$ 
and the supervised LASSO estimator with varying label size. In Scenario~\RomanNum{1} where the prior 
assumption holds exactly, PASS${}_{100}$, the PASS approach with $100$ labeled samples, even outperforms 
LASSO$_{400}$, the LASSO approach with $400$ labeled samples. When the prior assumption holds approximately
as in Scenarios \RomanNum{2} through \RomanNum{5}, PASS${}_{100}$ consistently outperforms
LASSO$_{150}$, and achieves similar performance as LASSO$_{200}$, which requires twice as many labels.
Finally, in Scenarios \RomanNum{6} where the prior information is highly inaccurate,
the PASS method maintains comparable performance against LASSO${}_{100}$.

\def\bgamma{\boldsymbol{\gamma}}
\def\bfeta{\boldsymbol{\eta}}

\subsection{Efficiency and Robustness Evaluations under Mis-specifications}\label{sec:sim:add}

{\purple
We conducted simulation studies under three additional scenarios to further investigate the efficiency and robustness of PASS when the model assumptions and elliptical design assumptions are violated. 
We again set $p=500$ and generate $\bX_{i}=2\Phi(\bZ_{i})-1$, where $\Phi(\cdot)$ is the cumulative distribution function of the standard normal, $\bZ_i=(Z_{i1},\ldots,Z_{ip})^\T\sim N(\zero, \bSigma_{\bZ}')$, $\bSigma'_{\bZ} = (\sigma'_{i,j})_{i, j = 1}^p$, $\sigma'_{i,j} = (0.5)^{\abs{i-j}}$ if $i=j$ or both $i$ and $j$ are $\leq 20$ or both $i$ and $j$ are $>20$ and $\sigma'_{i,j}=0$ otherwise. We make $\bSigma'_{\bZ}$ block-diagonal for the convenience of obtaining the population solution of $\bbeta$ and $\balpha$ through the best logistic or least square approximation under model mis-specifications. In real EHR studies, a paradigm of data generation is that the features $\bX$, e.g. some genetic variants, precedes the disease status $Y$, and $Y$ precedes some clinical surrogate $S$, e.g. the count of ICD codes associated with the disease. To mimic this scenario, we generated $Y_i$ and $S_i$ from the following models: 
\begin{alignat*}{2}
& Y_i=I\{ (0.8, 1, -1, 0.8, 0.4, \zero_{p-5}^\T)\bX_i+\epsilon_{yi} \ge 0\}, &\quad& \epsilon_{yi} \sim N(0,1), \\
& S_i=\mu Y_i+\bfeta_1\trans\bX_i+Y_i\bfeta_2\trans\bX_i + \epsilon_{si}, &\quad& \epsilon_{si} \sim N(0,1) .
\end{alignat*}
Assumptions $(\Cprior)$ and $(\ModelS)$ hold when $\bfeta_1=\bfeta_2=\zero$, and would be severely violated when $\bfeta_1$ and $\bfeta_2$ are large. We design three scenarios with $\bfeta_1$ and $\bfeta_2$ representing different degrees of violation on the surrogate assumptions:
\begin{itemize}
\item[i:] 
$\mu=1$, and $\bfeta_1=\bfeta_2=\zero$;
\item[ii:] 
$\mu=1.5$, $\bfeta_1=(\ba_3^\T,\zero_{p-5}^\T)^\T$, and $\bfeta_2=(\bd_3^\T,\zero_{p-5}^\T)^\T$;
\item[iii:] 
$\mu=2$, $\bfeta_1=(\ba_3^\T,\ba_3^\T,\ba_3^\T,\zero_{p-15}^\T)^\T$, and $\bfeta_2=(\bd_3^\T,\bd_3^\T,\bd_3^\T,\zero_{p-15}^\T)^\T$,
\end{itemize}
where $\ba_3=(0.6, -0.4, 0.4, 0.5, -0.5)^\T$ and $\bd_3=(0.3, 0.4, 0.6, -0.5, - 0.5)^\T$. Here $\mu$ depict the marginal effect of $Y_i$ on $S_i$, and are set to make the AUC of target models at a similar level across the three scenarios. Across all scenarios, $P(Y_i =1 \mid S_i,\bX_i)$ is no longer a parametric logistic model, i.e. $(\ModelY)$ is misspecified. Our goal is to estimate the limiting coefficients $\zeta_0,\gamma_0,\bbeta_0$ defined as the minimizor of $\E\ell(Y_i, \zeta + \gamma S_i +  \bX_i^\T \bbeta)$. Benchmark methods, and their implementation, tuning, and evaluation procedures are the same as in Section \ref{sec:sim:main}, except that we implement supervised LASSO with $n$ ranging from $100$ to $700$. 

\begin{figure}[htbp]
    \centering
    \caption{{\purple \label{fig:sim:mis} Evaluation metrics on the test set for simulation studies under Scenarios i--iii introduced in Section~\ref{sec:sim:add}.
Outliers are not drawn. Mean performance of the PASS approach are marked using red dash lines for ease of comparison. On the left panel, we present the evaluation metrics of all methods for comparison when $n=100$. On the right panel, we compare the performance of PASS when $n=100$ with supervised LASSO obtained using labelled samples with various $n$ (from $100$ to $700$).}}
\begin{minipage}{1\textwidth}\centering
\mbox{
    \includegraphics[width=0.48\textwidth]{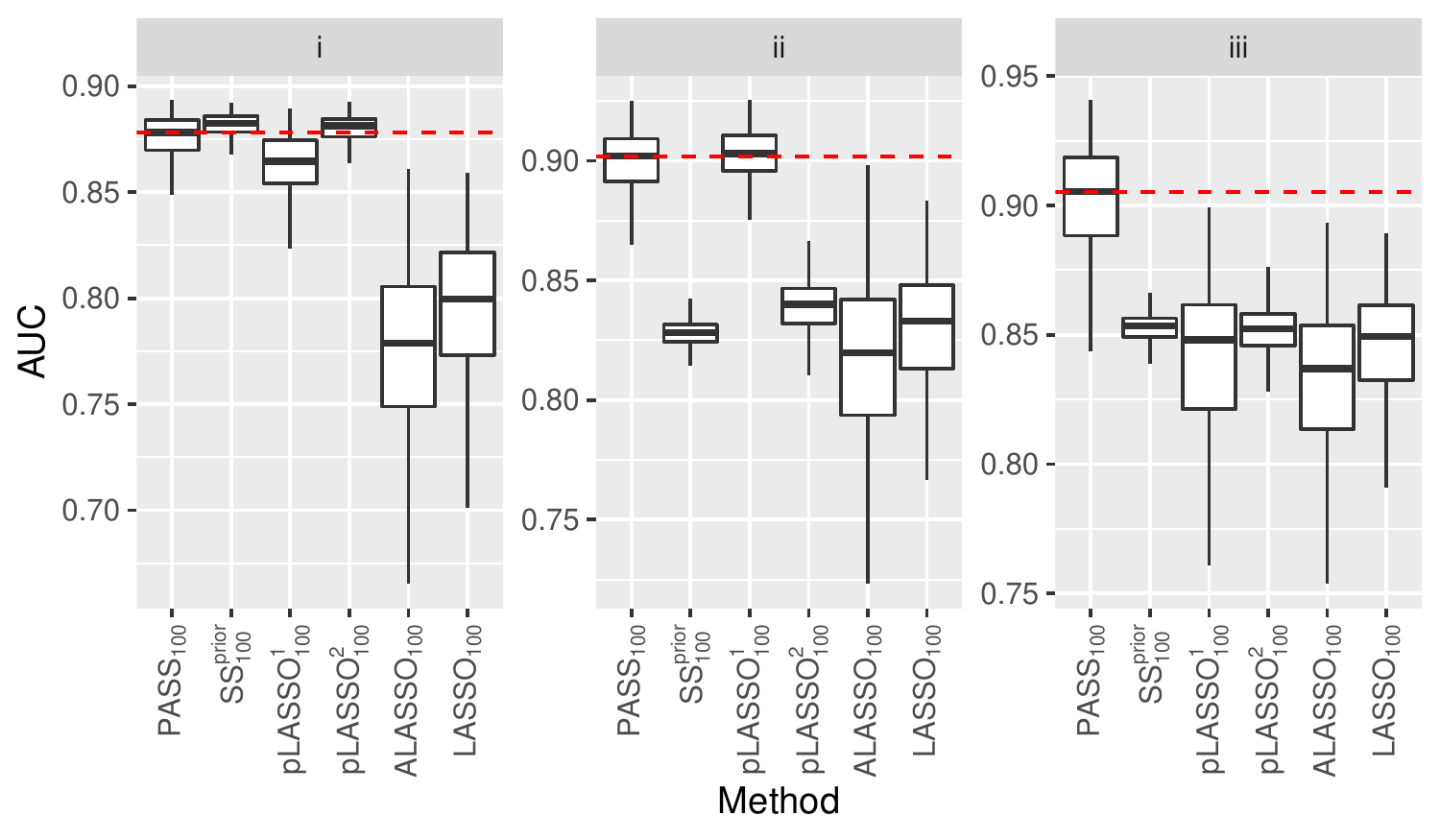}
    \includegraphics[width=0.48\textwidth]{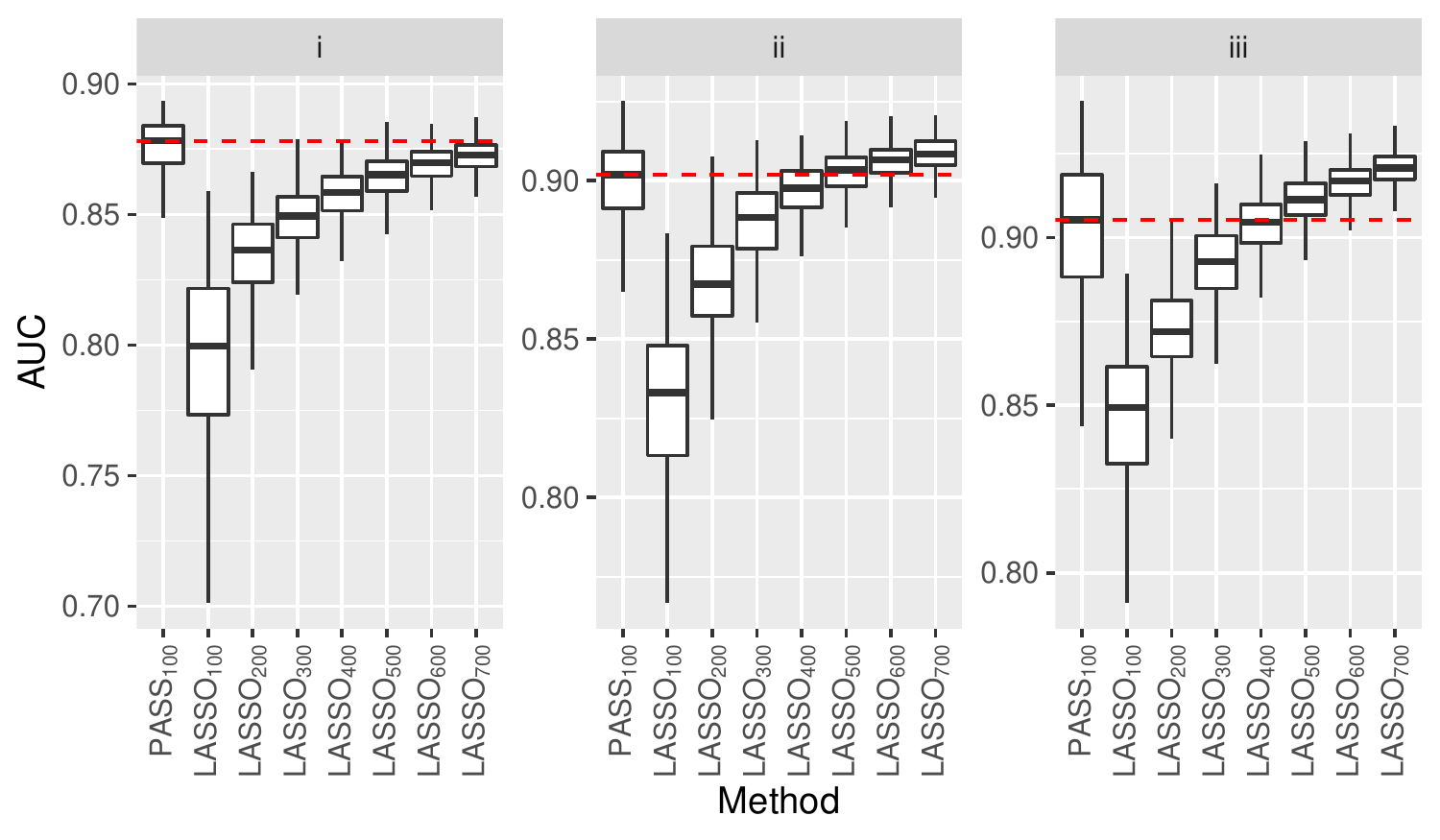}
}
\end{minipage}
\vspace{0.15in}
\begin{minipage}{1\textwidth}\centering
\mbox{
    \includegraphics[width=0.48\textwidth]{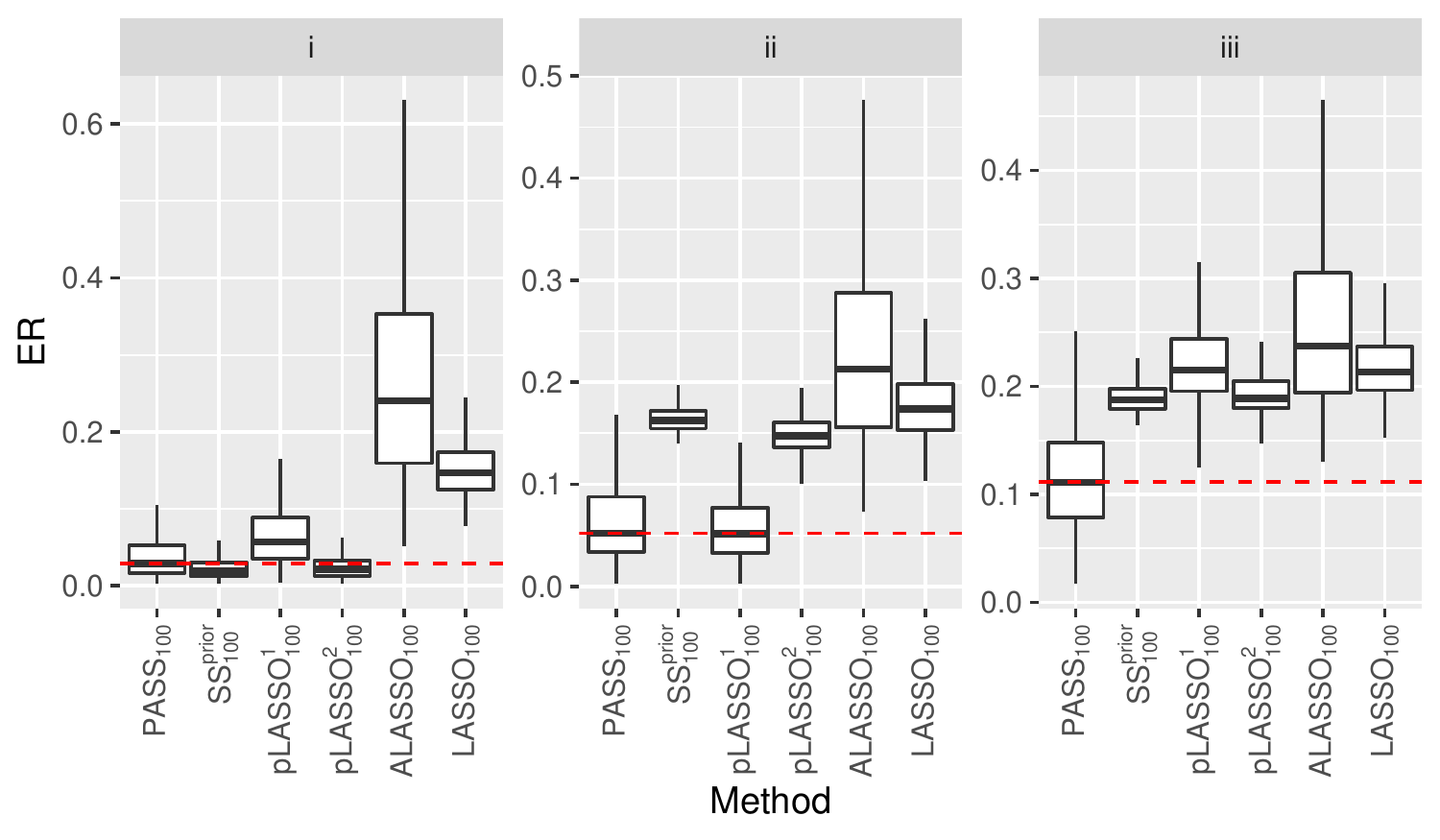}
    \includegraphics[width=0.48\textwidth]{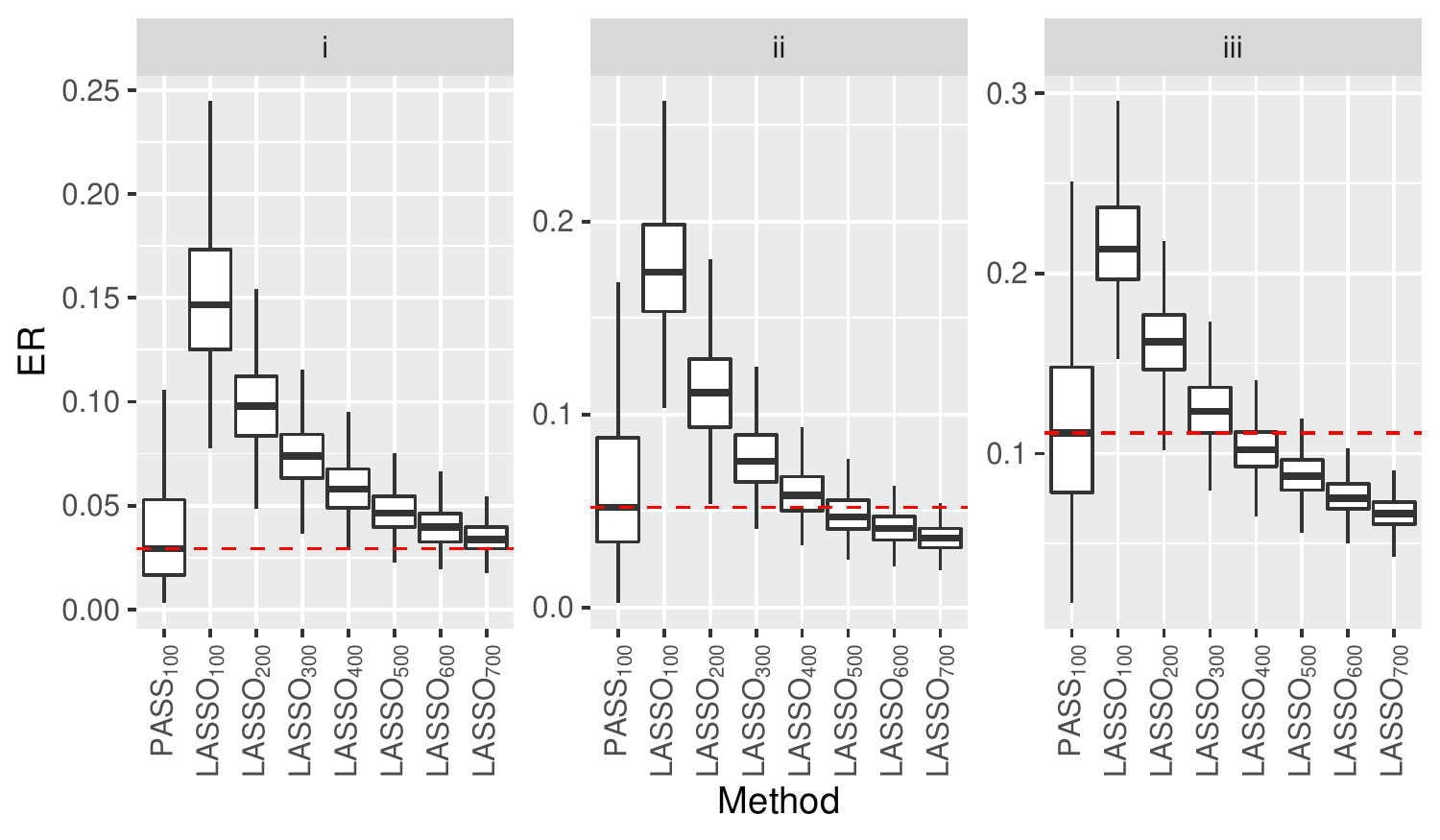}
}
\end{minipage}
\vspace{0.15in}
\begin{minipage}{1\textwidth}\centering
\mbox{
    \includegraphics[width=0.48\textwidth]{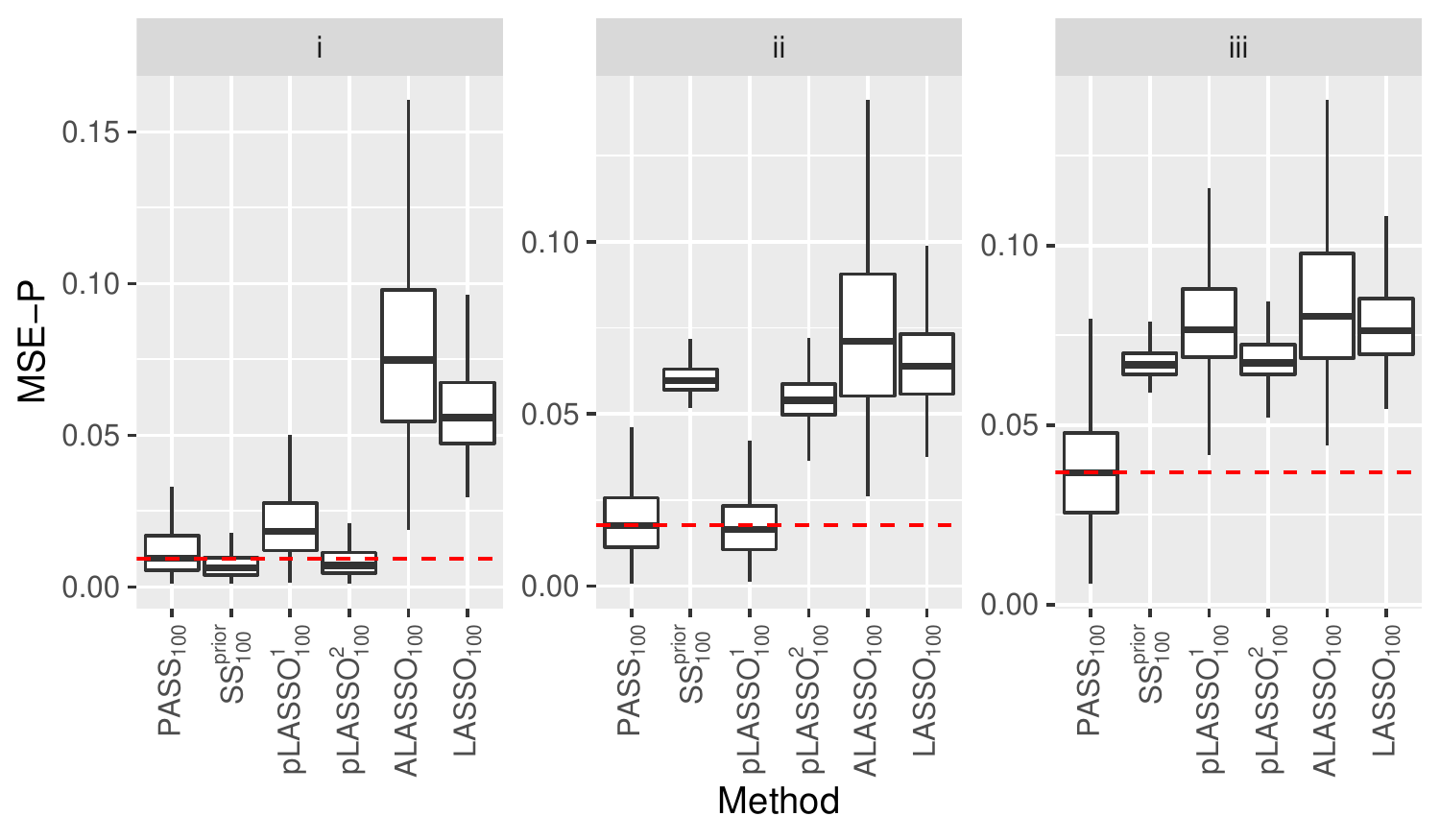}
    \includegraphics[width=0.48\textwidth]{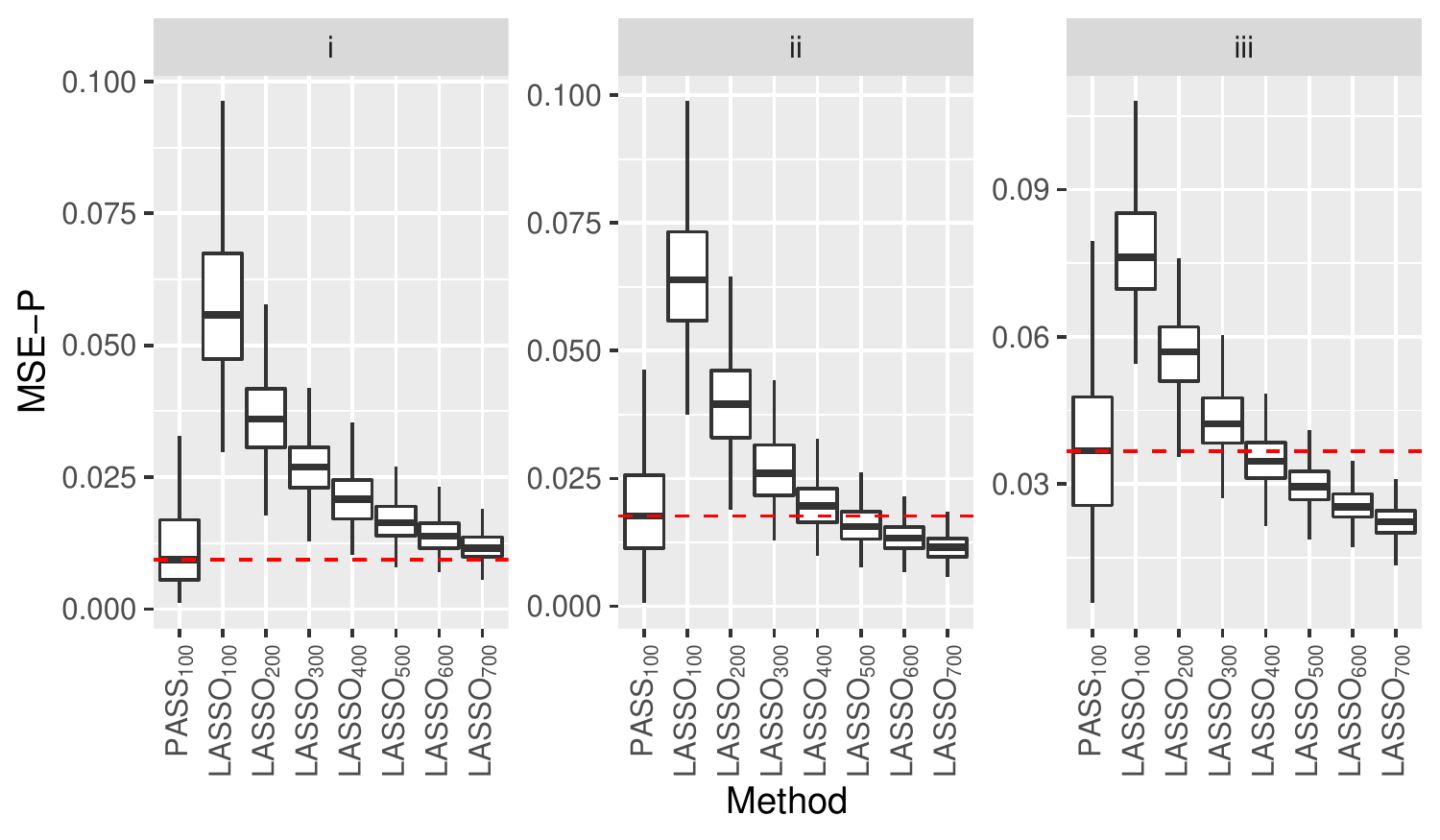}
}
\end{minipage}
\end{figure}

In Figure \ref{fig:sim:mis}, we present AUC, ER and MSE-P of the methods under Scenarios i--iii. In Scenario i, PASS has similar performances as the semi-supervised benchmarks \SSprior and pLASSO and all the semi-supervised estimators significantly outperform the two supervised estimators since $(\Cprior)$ holds and $\balphaast$ basically recover the direction of $\bbeta_0$ well. Among the semi-supervised estimators, the \SSprior and pLASSO$^2$ estimators have a slight advantage with smaller variation as expected since both heavily rely on the prior information which is of high quality in this setting. In Scenario ii, the key assumption $(\Cprior)$ is violated, which drastically impacts the performance of \SSprior and pLASSO$^2$. On the other hand, PASS and pLASSO$^1$ still effectively leverage the imperfect information from $\balphaast$ to approximately recover the support of $\bbeta_0$, and thus outperform \SSprior, pLASSO$^2$, and the supervised methods. In Scenario iii, $\bfeta_1$ and $\bfeta_2$ become denser than those in Scenario ii. This can make the recovery of $\supp(\bbeta_0)$ using $\supp(\balphaast)$ less accurate, and interestingly, PASS outperforms all methods including pLASSO$^1$ that also leverages $\supp(\balphaast)$. In all the three scenarios, PASS significantly outperforms supervised LASSO using the same or even 2-3 more times of sample sizes, which displays a large gain of using the unlabelled data to assist the regression. Finally, the results demonstrate that our method can still efficiently leverage the prior information from $S$ in estimating the target parameters when $S\mid Y,\bX$ highly depends on $\bX$ so $\Cprior$ is violated, ($\ModelY$) is misspecified, and the design is non-elliptical.

}

\def\sX{\mathcal{X}}
\def\sN{\mathcal{N}}

\section{Application to EHR Phenotying}\label{sec:apply}
{\purple
We examine the performance of PASS along with other approaches in three real world EHR phenotyping studies with the goal of developing classification models for the diseases of interest.  All studies are performed at a large tertiary hospital system with EHR data spanning over multiple decades. Each study has $n_0$ labeled observations for algorithm training and validation. We consider three choices of training size $n$ no more than $n_0/2$ in all examples. First, we randomly split the labelled samples into four folds of equal sizes. Then we pick each fold as the validation set, sample $n$ training labels from the other three folds for $20$ times, train and validate the algorithms, and finally average the evaluation metrics and their standard errors over the validation results on the four folds. We replicate this procedure for $10$ times and report the average performance.

\begin{rexample}[CAD Phenotyping]
The goal of this study is to identify patients with coronary artery disease (CAD) based on their EHR features. The study cohort consists of $N=4164$ patients, out of which a random subset of $n_0 = 181$ patients have their true CAD status annotated via chart review by domain experts. We use the sum of the counts for the CAD ICD code and NLP mention of CAD as the surrogate. There are $p=585$ additional EHR features consisting of the total count of all ICD codes as a healthcare utilization measure, $10$ ICD codes related to CAD, and $574$ NLP variables. For the size of training labels, we consider $n=50,70,90$. This de-identified dataset has been analyzed in previous studies \citep[e.g.]{zhang2019high} and is publicly available online: \url{https://celehs.github.io/PheCAP/articles/example2.html}.

\label{data:ex:1}
\end{rexample}

\begin{rexample}[RA Phenotyping]
Similar to the CAD phenotyping study, the goal is to identify patients with rheumatoid arthritis (RA) based on their EHR features. There are $N=46114$ patients in total and out of which, $n_0=435$ patients have their RA status annotated.  Again, we choose the sum of the ICD code and NLP mention of RA as the surrogate. The $p=924$ additional EHR features consist of the healthcare utilization and $923$ NLP variables potentially predictive of RA. For the size of training labels, we consider $n=50, 125, 200$.

\label{data:ex:2}
\end{rexample}

\begin{rexample}[Depression Phenotyping]
The goal is to identify patients with depression based on their codified EHR features. There are $N=9474$ patients in total and $n_0 = 236$ labeled observations. The surrogate is chosen as the counts of depression ICD code. There are $p=231$ additional EHR features, including the healthcare utilization and $230$ codified EHR features on depression related medication prescriptions, laboratory tests and ICD codes. For the size of training labels, we consider $50, 85, 120$.
\label{data:ex:3}
\end{rexample}

In the three data examples, $N$ is significantly larger than $p$ with $N/\max(p,n)$ being approximately $7$ for CAD, $50$ for RA, and $41$ for Depression. In all the three studies, we apply $x \to \log(1+x)$ transformation for all count variables. 
Also, since patients with higher healthcare utilization tend to 
have higher counts of most features, we orthogonalize all features against the healthcare utilization before regression fitting.} Since $\bvartheta_0$ is unknown in applications, we quantify the performance of an estimator $\wt\bvartheta$ based on the AUC and Brier skill score (BSS) of $\pi(\wt\bvartheta\trans\bW)$ for predicting $Y$, where the BSS is defined as $1-\wh \E_v[\{ Y-\pi(\wt\bvartheta\trans\bW)\}^2] / \wh\E_v[\{Y-\wh\E_v(Y)\}^2]$, and $\wh \E_v$ denotes the empirical expectation on the validation sample. The BSS is essentially a binary version of the R-square.

For comparison, we included PASS, \SSprior, pLASSO$^2$, supervised LASSO and ALASSO on the three data examples to estimate the phenotyping model $(\ModelY)$. {\purple We exclude  pLASSO$^1$ since it requires fitting of an unpenalized regression on $\supp(\balphahat)$, which is infeasible when $|\supp(\balphahat)|>n$. In addition, we compare to the unsupervised LASSO (ULASSO) approach of \cite{chakrabortty2017},  which estimates direction of the logistic coefficients for $Y\sim \pi(\bbeta\trans\bX)$ by regressing $I(S>c_u)$ against $\bX$ on the subset whose $S$ is either greater than $c_u$ or smaller than $c_l$, for some pre-specified $c_u$ and $c_l$ typically chosen such that $P(S > c_u)$ and $P(S < c_l)$ are small. Since the ULASSO approach only provides an estimate $\widetilde\bbeta$ to optimize the prediction of $\bbeta\trans\bX$ for $Y \mid \bX$ without using $S$ explicitly as an additional predictor, we also derive a semi-supervised variant of ULASSO, denoted by $\SSulasso$, by regressing the labeled $Y$ against $\widetilde\bbeta\trans\bX$ and $S$ as for  \SSprior. 




}





\begin{figure}[htbp]
\centering
\caption{\label{fig:example} {\purple Out of sample AUC and BSS on the data examples \ref{data:ex:1}--\ref{data:ex:3}, with various sizes of labelled training samples denoted as $n$. Median performance of PASS are marked using red dash line for ease of comparison.}} 
\subcaption{CAD}
\begin{minipage}{1\textwidth}\centering
\mbox{
\includegraphics[width=0.48\textwidth]{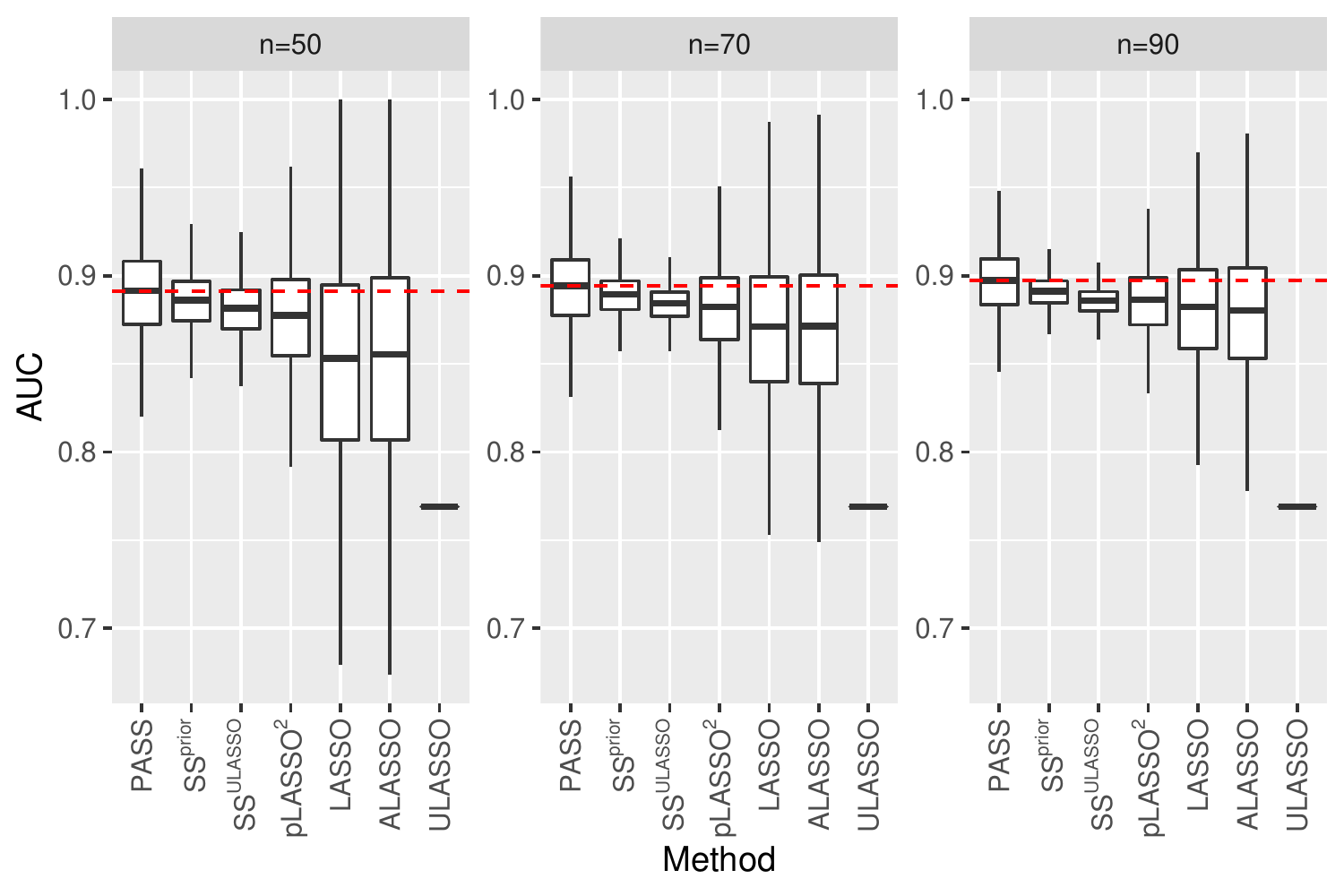}
\includegraphics[width=0.48\textwidth]{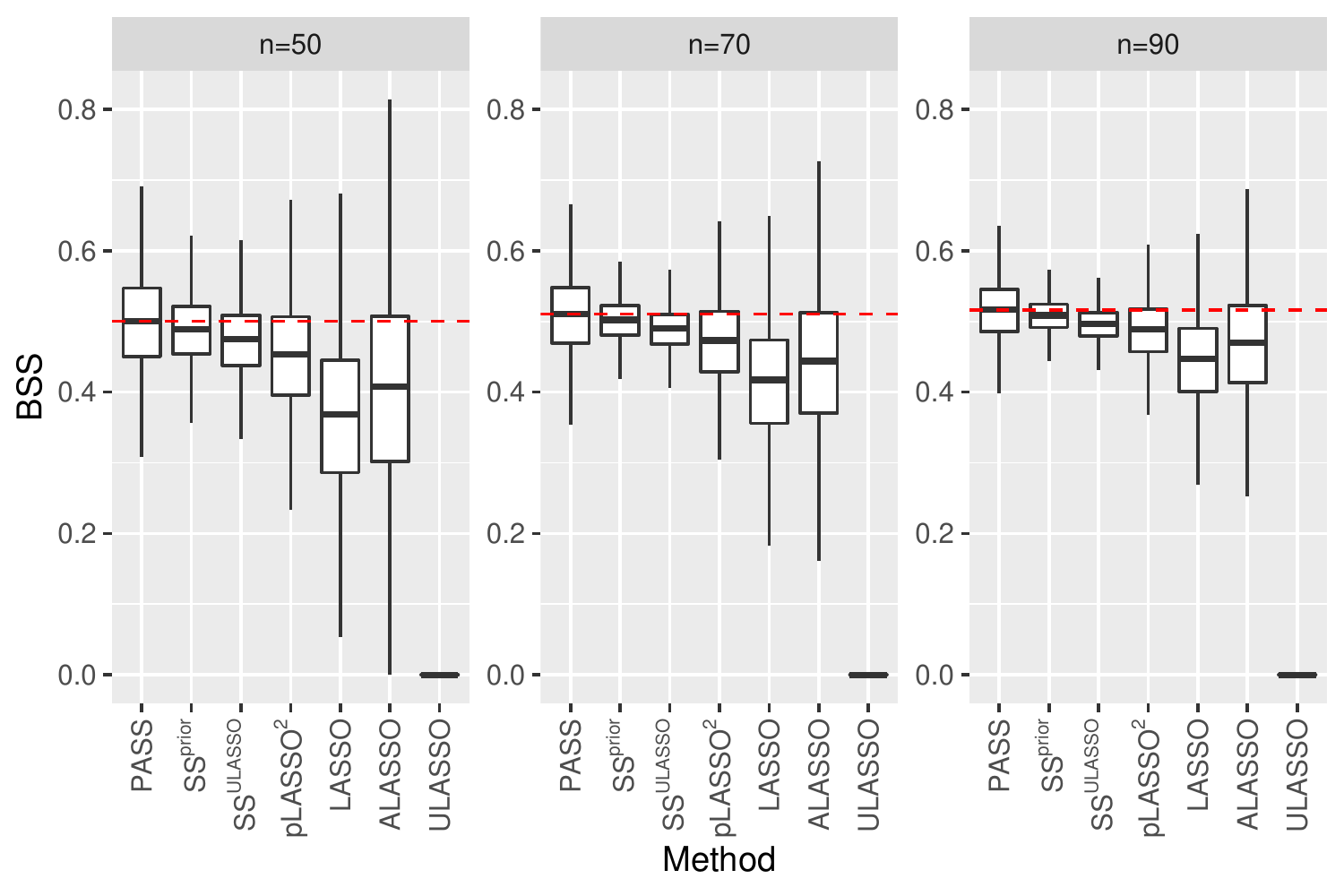}}
\end{minipage}
\vspace{0.12in}
\subcaption{RA}
\begin{minipage}{1\textwidth}\centering
\mbox{
\includegraphics[width=0.48\textwidth]{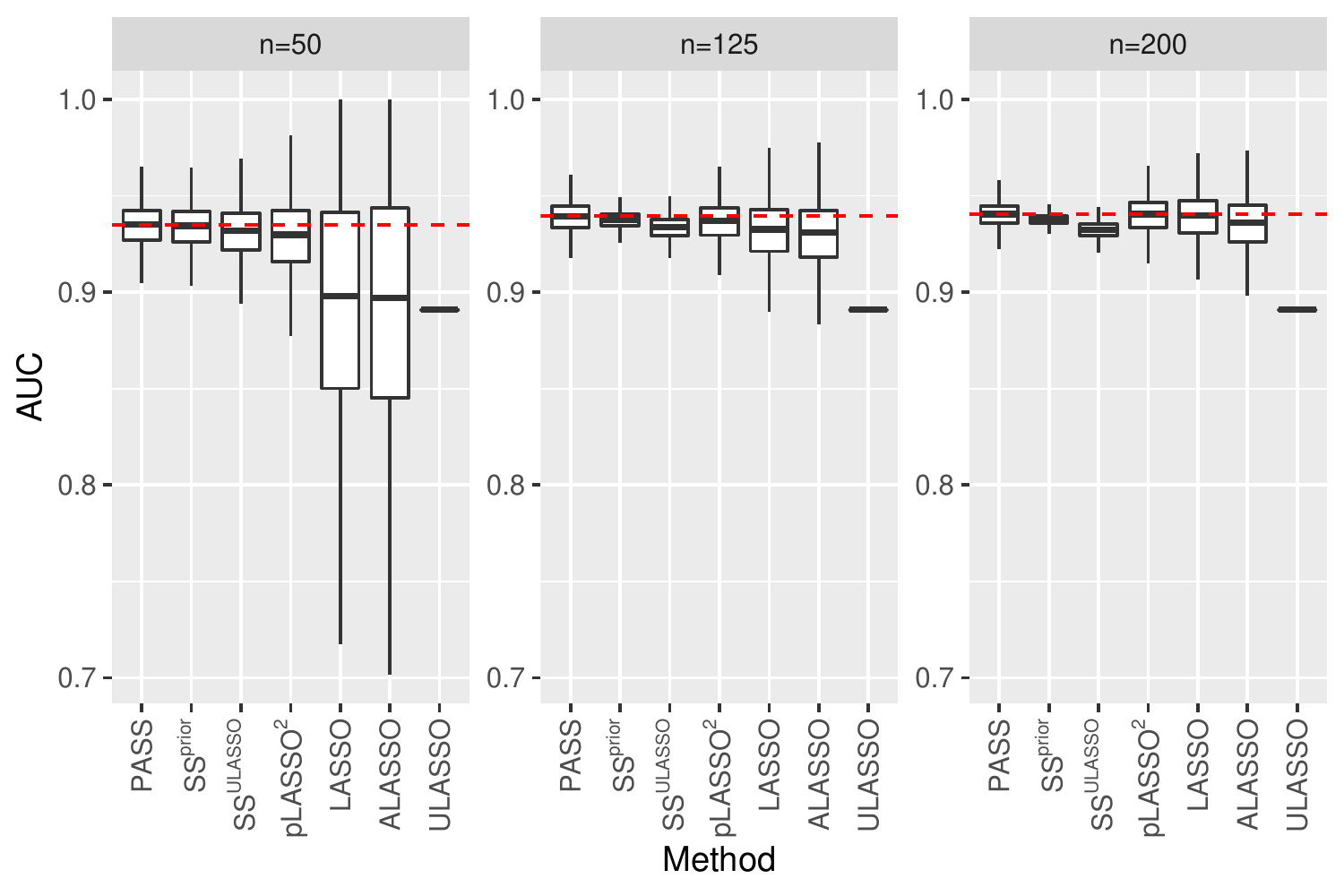}
\includegraphics[width=0.48\textwidth]{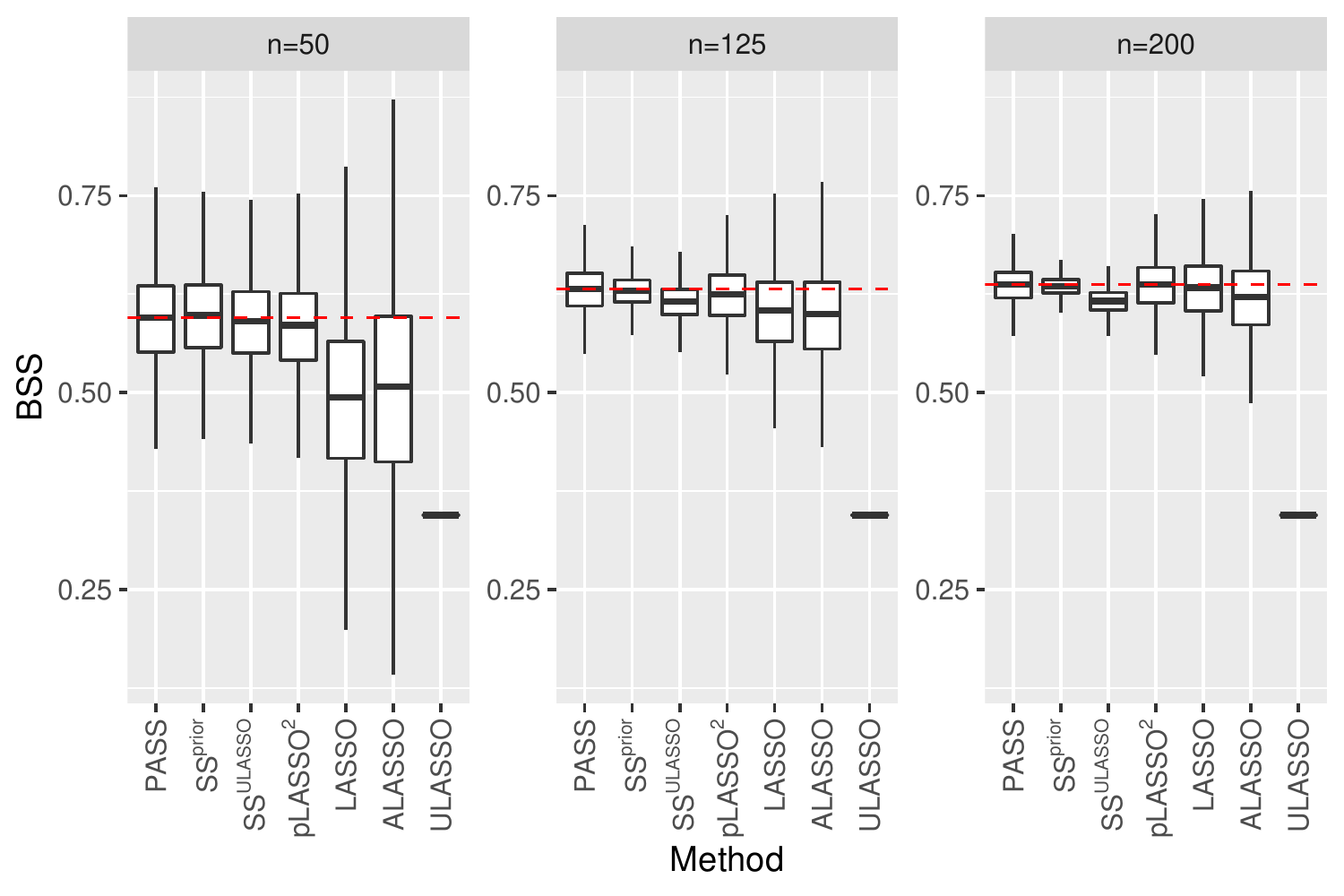}}
\end{minipage}
\vspace{0.12in}
\subcaption{Depression}
\begin{minipage}{1\textwidth}\centering
\mbox{
\includegraphics[width=0.48\textwidth]{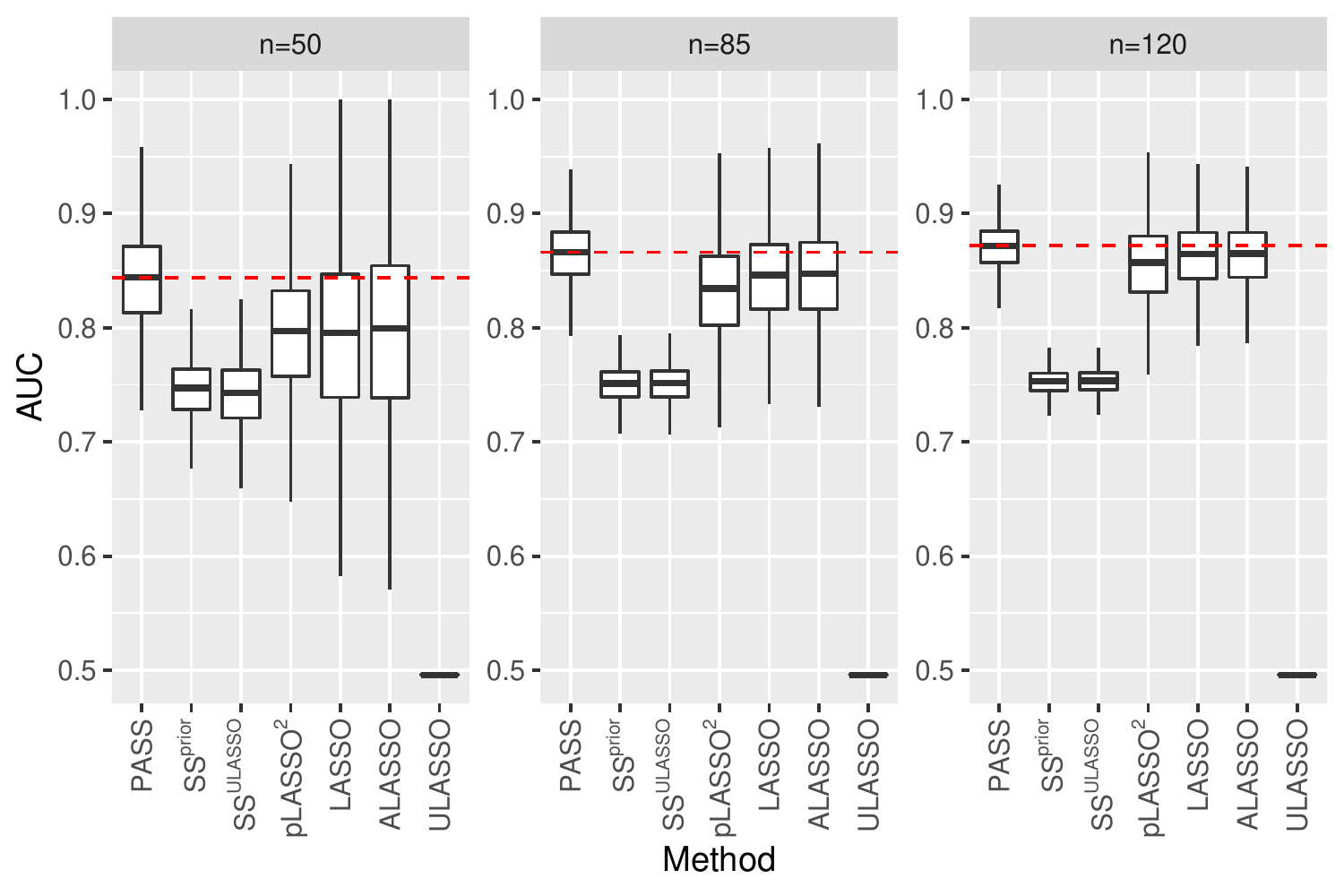}
\includegraphics[width=0.48\textwidth]{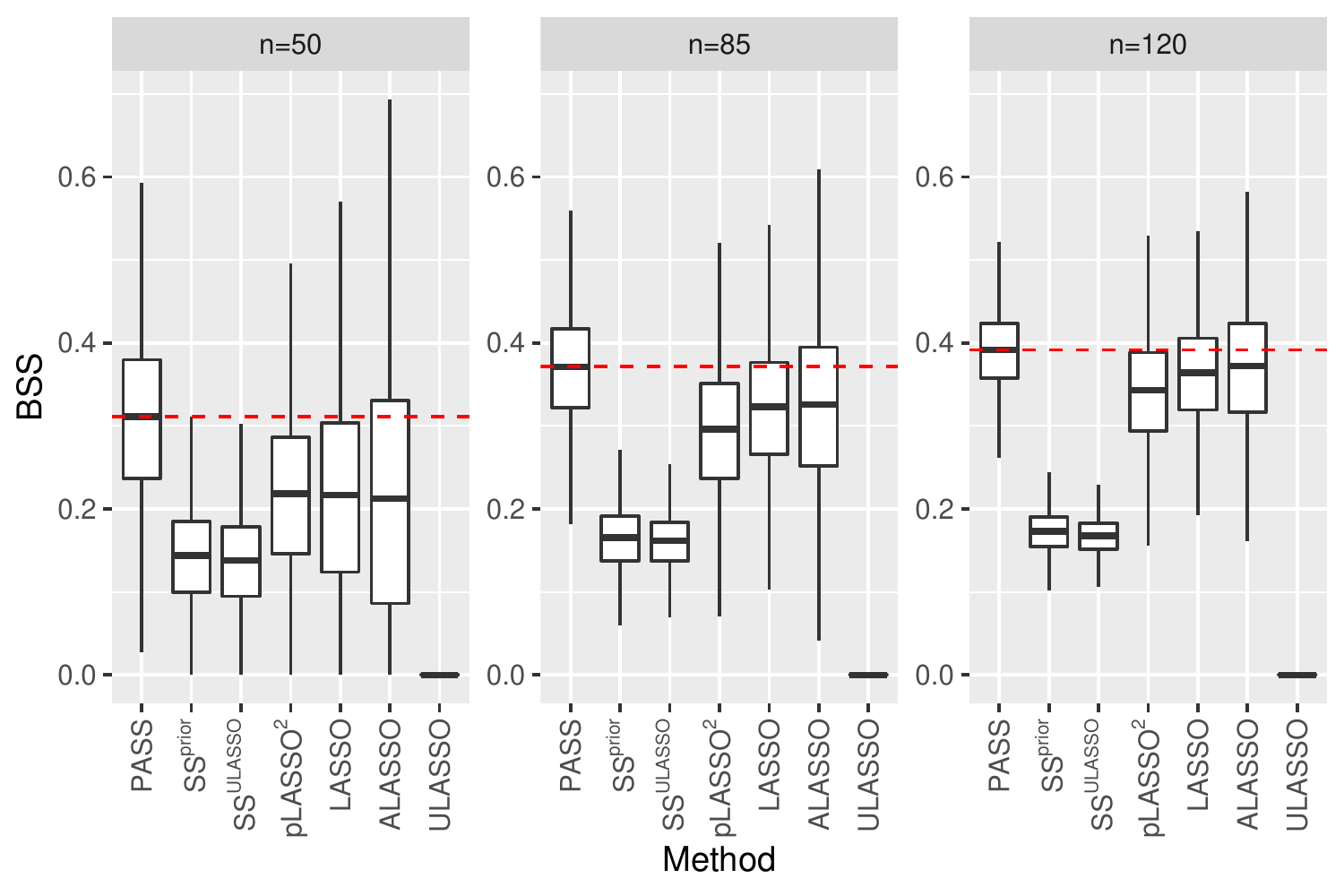}}
\end{minipage}

\end{figure}

{\purple
As shown in Figure \ref{fig:example}, PASS significantly outperforms the supervised LASSO and ALASSO when $n=50$ on all the three examples. As the label size $n$ increases, their performances get closer. Compared with the semi-supervised benchmarks, PASS has slightly or moderately better performance on the CAD and RA studies. For Depression, PASS substantially outperforms them, especially \SSprior and \SSulasso. For example, when $n=50$, PASS attained average AUC in classifying depression about $0.1$ higher than that of \SSprior and \SSulasso and $0.05$ higher than pLASSO. The gap becomes smaller when $n$ increases as  expected. Interestingly, the supervised estimators outperform pLASSO, \SSprior, and \SSulasso on the Depression data as well but has similar or worse performance than these semi-supervised approaches on the other two examples. This could in part be attributed to the relatively poor quality of the surrogate information, which makes existing semi-supervised approaches fail. In contrast, PASS could utilize such prior information more effectively and robustly, and still preserves better performance than the supervised estimators. Thus, we can conclude that incorporating prior information from the unlabeled data could improve and stabilize the prediction performance of phenotyping models in EHR applications, and PASS is more robust and efficient in leveraging the prior information compared with existing semi-supervised methods. In addition, ULASSO shows much worse performance than the other supervised and semi-supervised methods on all examples. This illustrates the importance of collecting labels and including the surrogate in the regression models for EHR phenotyping.


}

\section{Discussion}

In this paper, we propose PASS, a high dimensional sparse estimator adaptively incorporating the prior knowledge from surrogate under a semi-supervised scenario commonly found in application fields like EHR analysis. {\purple Compared to the supervised approaches, the proposed PASS approach can substantially reduce the required number of labeled samples when the model assumptions ($\ModelS$) and $(\Cprior)$ and the elliptical design assumption (C1) hold exactly or approximately, and thus the prior information $\balpha^*$ is trustworthy. Compared to existing pLASSO and $\SSprior$ approaches that also incorporates prior information, the PASS approach is robust against unreliable prior information $\balpha^*$, which might be the case when the surrogate model assumptions are violated or the design $\bX$ is highly non-elliptically distributed.}

One of the main challenges in our theoretical analysis comes from the colinearity of covariates $(1, S_i, \bX_i^\T \balphahat, \bX_i^\T)^\T$ due to the enrollment of $\rho$ to leverage the prior information in $\balphahat$. We overcome this by properly constructing the oracle coefficients $\bthetaast$ and the restricted eigenvalue assumption (A6). The formulation of our problem falls into the missing data framework with missing completely at random. However, the missing
probability approaches 1 as $N \to \infty$. This together with the high dimensionality of $\bX$ makes the theoretical justifications
more challenging than those used in the standard missing data literature. Without prior assumptions of $\bbeta_0 - \rho \balpha_0$
being sparse in certain sense, the unlabeled data cannot directly contribute to the estimation of $\bbeta_0$. Our proposed PASS
procedure hinges on the sparsity of $\bbeta_0- \rho \balpha_0$ to leverage the unlabeled data.

\def\subone{^{\sf \scriptscriptstyle [1]}}
\def\supk{^{\sf \scriptscriptstyle [k]}}
\def\supK{^{\sf \scriptscriptstyle [K]}}

We have restricted the discussion to a single surrogate variable for simplicity. However, the proposed method 
can be easily extended to multiple surrogates. Specifically, consider $K$ surrogates, denoted by $S\subone, 
\dots, S\supK$. Let $\balphahat\supk$ be the ALASSO estimator regressing $S_{i}\supk$ against $\bX_i$,
$\cAhat = \cup_{k=1}^K \supp(\balphahat_k)$, $\bm{S}_i = (S_{i}\subone, \dots, S_{i}\supK)^\T$
and $\bm{\rho} = (\rho\subone, \dots, \rho\supK)^\T$.
We can obtain an estimator for the model parameters as
\[
\zetahat, \wh{\bm{\gamma}}, \wh{\bm{\rho}}, \bbetahat = 
\argmin_{\zeta, \bm{\gamma}, \bm{\rho}, \bbeta} 
\ninv \sum_{i=1}^n
\ell(Y_i, \zeta + \bm{S}_i^\T \bm{\gamma} + \bX_i^\T \bbeta) + 
\lambda_1 \norm[1]{(\bbeta - \sum\nolimits_k \rho_k \balphahat_k)_{\cAhat}} +
\lambda_2 \norm[1]{\bbeta_{\cAhat^c}}.
\]
Theoretical justification and finite sample performance of $\bbetahat$ under this setting warrant further research. {\purple In addition, it may be interesting to extend the semi-supervised PASS estimator under a high dimensional sparse parametric regression to semi-parametric settings such as sparse additive model \citep{ravikumar2009sparse} and sparse varying coefficient model \citep{noh2010sparse}. Under semi-parametric models, one could still leverage prior information through shrinking the coefficients to ``$\rho\balphahat$" with some sparse penalty function, to gain statistical efficiency. Studying the specific forms and theoretical properties of such approaches via a  semi-supervised framework warrants future research.


{\bf R} codes for implementing PASS and the benchmark methods, and replicating the simulation results can be found at \url{https://github.com/moleibobliu/PASS}. 
}

\bibliographystyle{apalike}
\bibliography{highdim}

\end{document}